\documentclass[twocolumn,pre,superscriptaddress,nofootinbib]{revtex4}
\usepackage{tikz}
\usepackage{graphicx}%
\usepackage{color} 
\usepackage{transparent}
\usepackage{dcolumn}
\usepackage{amsmath}
\usepackage{amssymb, bm, wasysym}
\usepackage{latexsym}
\usepackage{hyperref} 
\usepackage{booktabs}
\usepackage{latexsym, bbold}
\begin{document}

\def\bea{\begin{eqnarray}}
\def\eea{\end{eqnarray}}
\def\beq{\begin{equation}}
\def\eeq{\end{equation}}
\def\f{\frac}
\def\k{\kappa}
\def\e{\epsilon}
\def\ve{\varepsilon}
\def\be{\Pi r_0^3}
\def\D{\Delta}
\def\h{\theta}
\def\t{\tau}
\def\a{\alpha}
\def\le{\left}
\def\ri{\right}

\def\ve{\varepsilon}
\def\fv{\vec{f}}
\def\fm{\vec{f}_m}
\def\zh{\hat{z}}
\def\yh{\hat{y}}
\def\xh{\hat{x}}
\def\km{k_{m}}

\def\cDa{d\varsigma}
\def\cD{{\cal D}[x]}
\def\cL{{\cal L}}
\def\cLo{{\cal L}_0}
\def\cLa{{\cal L}_1}

\def\Re{{\rm Re}}
\def\sj{\sum_{j=1}^2}
\def\rk{\rho^{ (k) }}
\def\rek{\rho^{ (1) }}
\def\cek{C^{ (1) }}
\def\rz{\rho^{ (0) }}
\def\rt{\rho^{ (2) }}
\def\rtb{\bar \rho^{ (2) }}
\def\trk{\tilde\rho^{ (k) }}
\def\trek{\tilde\rho^{ (1) }}
\def\trz{\tilde\rho^{ (0) }}
\def\trt{\tilde\rho^{ (2) }}
\def\r{\rho}
\def\tD{\tilde {D}}

\def\s{\sigma }
\def\kb{k_B}
\def\la{\langle}
\def\ra{\rangle}
\def\nn{\nonumber}
\def\up{\uparrow}
\def\dn{\downarrow}
\def\S{\Sigma}
\def\dg{\dagger}
\def\d{\delta}
\def\p{\partial}
\def\l{\lambda}
\def\L{\Lambda}
\def\G{\Gamma}
\def\o{\Omega}
\def\w{\omega}
\def\g{\gamma}

\def\noi{\noindent}
\def\a{\alpha}
\def\d{\delta}
\def\p{\partial} 

\def\la{\langle}
\def\ra{\rangle}
\def\e{\epsilon}
\def\n{\eta}
\def\g{\gamma}
\def\rv{{\bf r}}
\def\tv{{\bf t}}
\def\on{{\omega_{\rm on}}}
\def\off{{\omega_{\rm off}}}
\def\hf{\frac{1}{2}}


\title{Pattern formation, localized and running pulsation on active spherical membranes}

\author{Subhadip Ghosh}
\email{sghosh@phy.hr}
\affiliation{Department of Physics, Faculty of Science, University of Zagreb, Bijeni{\^c}ka cesta 32, 10000 Zagreb, Croatia}

\author{Sashideep Gutti}
\email{sashideep@hyderabad.bits-pilani.ac.in}
\affiliation{Department of Physics, BITS Pilani Hyderabad Campus, Hyderabad 500078, Telengana, India}

\author{Debasish Chaudhuri}
\email{debc@iopb.res.in}
\affiliation{Institute of Physics, Sachivalaya Marg, Bhubaneswar 751005, India}
\affiliation{Homi Bhaba National Institute, Anushaktigar, Mumbai 400094, India}

\date{\today}

\begin{abstract}
Active force generation by actin-myosin cortex coupled to the cell membrane allows the cell to deform, respond to the environment, and mediate cell motility and division. Several membrane-bound activator proteins move along it and couple to the membrane curvature. Besides, they can act as nucleating sites for the growth of filamentous actin. Actin polymerization can generate a local outward push on the membrane.  Inward pull from the contractile actomyosin cortex can propagate along the membrane via actin filaments. We use coupled evolution of fields to  perform linear stability analysis and numerical calculations. As activity overcomes the stabilizing factors such as surface tension and bending rigidity, the spherical membrane shows instability towards pattern formation, localized pulsation, and running pulsation between poles.  We present our results in terms of phase diagrams and evolutions of the coupled fields.  They have relevance for living cells and can be verified in experiments on artificial cell-like constructs.
\end{abstract}

\maketitle


\section{Introduction}
The cell's ability to change its shape is crucial for many of its functions, e.g., cell motility and  division~\cite{Alberts2009}.  
The associated membrane deformations involve several physical forces~\cite{Kozlov2016}. 
For example, coupling of local membrane curvature to protein domains, such as the BAR-domain proteins, can sculpt membrane shapes~\cite{Mills2004, McMahon2005}. 
This mechanism alone can generate supramolecular organization of membrane-associated proteins, as has been shown recently for coccal bacteria {\em S. aureus}~\cite{Garcia-Lara2015, Agudo-Canalejo2017}. 
In eukaryotes, actin polymerization is stimulated locally on the membrane by molecular complexes like Arp2/3 complex and WASP family activators~\cite{Pantaloni2001, Doherty2008}. The inward growth of F-actins, with a rate controlled by cofilin and profilin, can generate active outward push on the membrane~\cite{Marcy2004, Kuhn2005, Gov2006}.  
The Arp2/3  branching agents generate a dynamic actin network beneath the cell membrane. The myosin-mediated contractility in the cytoskeleton can propagate to the membrane via this network~\cite{Medeiros2006, Doherty2008, Murrell2015}.  
In a highly branched part of the actin network such an active pull can affect a relatively wider region of  the membrane, in contrast, when the filaments are bundled the myosin pull transmits to highly localized parts of the membrane~\cite{Svitkina2003}.
The same actin network can generate both pushing and pulling forces on the membrane.  
Membrane ruffles and traveling waves on cell membranes are ubiquitous in many cell types~\cite{Allard2012,Ryan2012,Dobereiner2006}. 
The organization of proteins and lipids in multicomponent membranes, formation of membrane domains, and their interaction with the underlying cytoskeleton play important functional role in cell biology~\cite{Julicher1993, Julicher1996, Sankararaman2002, Hu2011, Banerjee2018, Mayor2004, Jacobson2007}. 
The theoretical description of active surfaces and their application in cell membranes, organelles, and epithelial tissues have seen significant advancement in recent years~\cite{Salbreux2017a, Morris2019, Mietke2019b, Mietke2019a, Maitra2014, Ramakrishnan2014b, Ramakrishnan2015a, Loubet2012, Turlier2019, Sadhu2019, Sadhu2018}.  
The combination of curvature sensing proteins and  transverse active forces due to membrane pumps or cytoskeleton showed possibilities of instability and  traveling waves on  open flat membranes~\cite{Prost1998, Ramaswamy2000, Shlomovitz2007, Veksler2007, Shlomovitz2008, Chen2009a, Gov2009, Alaoui2009, Risler2015, Duclut2019}. 
On the other hand,  on closed membranes, the emergent features can depend on the superposition of deformation modes. 

In this paper, we consider active shape deformations of spherical membranes in the presence of curvature sensing activator proteins~(AP) and the cytoskeleton.  Many of the immune system cells, including lymphocytes, neutrophils, basophils, are naturally spherical. With the loss of cell adhesion and near cell division, isolated cells adopt a spherical shape~\cite{Stewart2011}.  
They can maintain their volume by osmotic control and display surface tension regulation~\cite{Morris2001a}. 
In the bottom-up synthetic biology approach, cell-sized vesicles, liposomes, and emulsion droplets encapsulating cytoskeletal extracts and other proteins can constitute simple models of  artificial cells~\cite{Simon2019, Durre2018, Litschel2018a, Bashirzadeh2019,  Tsai2011, Carvalho2013, Keber2014, Fanalista2019, Litschel2021, Litschel2021a, Liu2009a}. 
Equilibrium properties of spherical membranes with volume constraints are controlled by their surface tension and bending rigidity~\cite{Canham1970, Helfrich1973, Seifert1997}. 
We consider membrane-cortex adhesion~\cite{Alert2015,Agudo-Canalejo2017}  
 that can be mediated by several ligand and receptor molecules, like Talin and ERM proteins~\cite{Tsujioka2012, Tsukita1999}.  
The adhesion strength depends on the molecular type and concentration of the ligands and receptors~\cite{Alert2015}.
The passive coupling of APs with local membrane curvature preferring {\em hills} or {\em valleys} can bring APs together to deform the membrane~\cite{Ramaswamy2000}. 
In addition, the active push due to actin polymerization supports hill formation. The contractile pull by motor proteins from within the cytoskeleton can generate local valleys. 
In this paper, we explore the resultant coupled dynamics using linear stability analysis and numerical calculations.

Depending on control parameters, the system shows three different non-equilibrium phases, characterized by instability towards pattern formation, localized pulsation, and running pulsation. 
The spontaneous localization of valley-forming APs coupled with a cytoskeletal pull can lead to a deep inward deformation. 
Reduction of stabilizing properties like surface tension can assist such membrane instabilities.
The competition between hill and valley forming activities coupled to a feedback mechanism and phase lag in the dynamics can induce pulsation on the membrane.
Depending on parameter values, we observe traveling pulsation similar to that in open and flat membranes. 
On a closed membrane, in contrast, counter-propagating waves can superpose to form standing waves. We find such localized pulsations on the spherical membrane over a broad parameter range.  
 A living cell may exploit these properties in producing intended cell shapes, e.g., during cell migration and  cell division. On the other hand, the mechanisms we study can be incorporated into artificial cells, making our  predictions amenable to direct experimental verifications.

In Sec.~\ref{model}, we present the continuum model for the membrane coupled to APs, the impact of non-equilibrium processes, and the differential equations describing the coupled dynamics. In Sec.~\ref{results}, we first perform linear stability analysis to demonstrate instability and the presence of unstable spirals. We further differentiate the formation of localized and running pulsations and the associated membrane deformations using particular examples.  Finally, in Sec.~\ref{discussion}, we conclude by presenting a discussion and  outlook. 


\section{Model} 
\label{model}
We consider deformations of a spherical fluid membrane coupled to activator proteins~(AP), driven by growing actin filaments and a contractile cytoskeleton. The membrane's mechanical properties are characterized by surface tension, bending rigidity, and membrane-cortex adhesion. In addition, an osmotic regulation can maintain the cell volume.   
The APs diffuse on the membrane, deforming the membrane and localizing depending on the membrane curvature. 
Moreover, they act as nucleation sites of actin filaments that polymerize inwards to generate local outward push on the membrane. In addition, the F-actins are pulled inward actively by the cortical layer of the cytoskeleton. 

\subsection{Equilibrium description of the membrane and APs}
The bending energy of the membrane can be expressed in terms of the Canham-Helfrich free energy~\cite{Canham1970, Helfrich1973} 
\bea
F_b = \hf \k \int ds\, (2 c_m - c_s)^2, \nn
\eea
where, $c_m$ denotes the local mean curvature, $c_s$ is the spontaneous curvature of the undeformed sphere, and $ds$ denotes the area element. Here $\k$ is the bending rigidity of the membrane. 
Here we neglect the Gaussian curvature term. In a multicomponent vesicle it  does not impact the vesicle morphology when different membrane patches have the same Gaussian curvature modulus~\cite{Julicher1993, Julicher1996, Hu2011}. Further, we restrict ourselves to  deformations that do not change the overall  topology.

We consider a membrane area reservoir maintaining a finite surface tension $\s$ leading to an energy cost
$$F_S = \s \int ds.$$ 
We assume the membrane to be tethered to a rigid cell cortex via an effective spring constant per unit area $K_t$~\cite{Alert2015, Agudo-Canalejo2017}. The strength of $K_t$ depends on possible direct coupling of the membrane and cortex via proteins, or non-specific interactions like steric repulsion, van der Waals attraction or electrostatic interactions.  

The local concentration of APs changes the bending energy cost by~(Appendix-\ref{appendix_theory})
\bea
F_{AP} = \hf \L \int ds\, \psi^2  + \k \bar H \int ds\, (2c_m - c_s)\psi, \nn
\eea 
where, $\psi$ denotes the  AP concentration per unit area, $\L$ is the inverse compressibility, and 
the coupling constant $\bar H$ controls whether the APs promote and prefer local {\em hill} ($\bar H < 0$) or {\em valley} ($\bar H >0$) on the membrane~(see Fig.\ref{fig:mod_cartoon}($a$)\,). A deformed spherical membrane along with its coupling to various force generators are indicated in  Fig.\ref{fig:mod_cartoon}($b$). The cell can regulate its volume, e.g., by pumping fluid in and out. We use a volume constraint 
such that $\d V = (r_0^3/3) \int d\o  [(1+u)^3 -1] = 0$.  
The total number of membrane-bound APs is assumed to be constant $N=\psi_0 \, 4\pi r_0^2 = r_0^2 \int d\o~ \psi $. In the following, we describe each of these terms in further detail.

The small relative deformations $u(\theta,\phi,t)$ around an unperturbed sphere of radius $r_0$, describes the coordinates on the deformed sphere $r(\theta,\phi,t) = r_0[1+u(\theta,\phi,t)]$ in the spherical polar coordinates.  
Keeping up to quadratic order in $u$, the bending energy can be expressed as (Appendix~\ref{appendixA})
\beq
F_b=2\kappa\int d\Omega\left[ u + \f{1}{2} \Delta_2 u \right]^2
\label{bendingenergy3}
\eeq
integrated over the solid angle $d\Omega = \sin\h d\h d\phi$. 
We use two-dimensional gradient 
$\nabla_2$ 
and Laplace-Beltrami operator  $\Delta_2$ 
on the unit sphere~(Appendix~\ref{appendixA}).

The change in area element is $ds = r_0^2\left(2u+u^2+\frac{1}{2} (\nabla_2u)^2 \right) d\Omega$. 
The volume constraint within quadratic order in $u$ gives the relation $\int u~d\o = -\int u^2~d\o$. Using this we can rewrite $ds=r_0^2\left(-u^2+\frac{1}{2} (\nabla_2u)^2 \right) d\Omega$. 
Thus the stretching energy term
\bea
F_S = \s r_0^2 \int \left[ - u^2+\frac{1}{2}(\nabla_2u)^2 \right] d\Omega. 
\eea

The energy cost due to tethering of the membrane to the actomyosin cortex~\cite{Alert2015,Agudo-Canalejo2017} can be expressed in terms of a harmonic potential
\bea
F_{t}= \f{K_t r_0^4}{2} \int d\Omega\, u^2,
\eea
where, $K_t$ denotes an effective spring constant per unit area. Its value increases with the density and spring constant of bound linkers between the membrane and cortex~\cite{Alert2015}~(Fig.\ref{fig:mod_cartoon}($b$)\,).

%
Keeping up to bilinear order in $u$ and $\psi$, and using $c_s = 2/r_0$, the free energy contributions from APs can be expressed as
\beq
F_{\rm AP} = \int d\Omega \le[ \hf \L r_0^2 \, \psi^2 - \k \bar H r_0\,\, \psi \le(  u + \hf \D_2 u \ri) \ri] .
\eeq
The total free energy of the coupled fields of the membrane deforming activator proteins $\psi(\o)$ and the deformation $u(\o)$ is
\beq 
F=F_b+F_{s}+ F_t + F_{\rm AP} . 
\label{tot_en}
\eeq
This describes the equilibrium morphologies of spherical membranes~\cite{Agudo-Canalejo2017} and in the limit of large $r_0$ reproduces known results for flat membranes~\cite{Ramaswamy2000}~(see Appendix-\ref{equi_mem}).

\subsection{Active forces}
\begin{figure}[!t]
\begin{center}
\includegraphics[width=8.6cm]{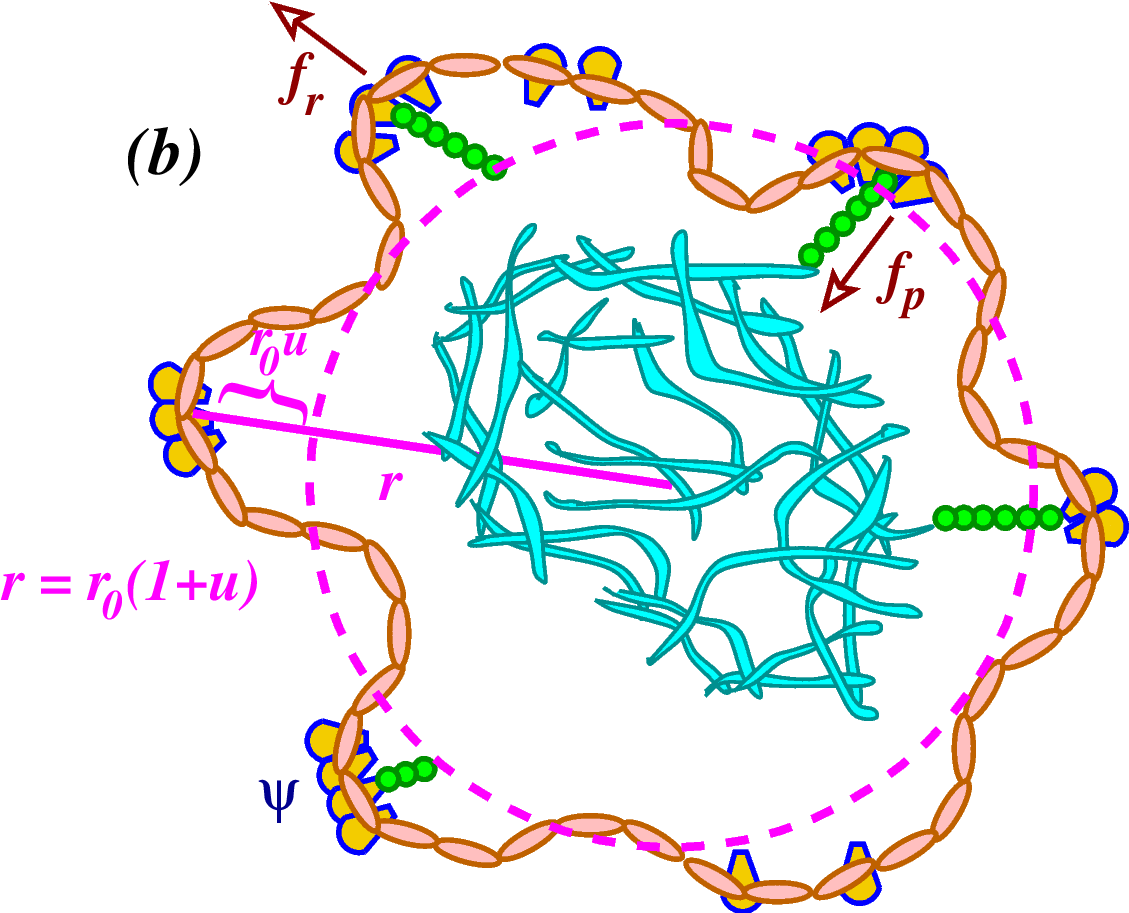}
\end{center}
\caption{(${\bm a}$)~Schematic diagram showing the significance of geometric shape of APs leading to inward ($\bar{H}\textgreater 0$) and outward ($\bar{H}\textless 0$) bulging of the membrane. (${\bm b}$)~Schematic diagram showing a cross section of the spherical membrane and essential components. $u$ and $\psi$ are local deformation of the membrane and local AP density respectively. $f_r$ and $f_p$ are active rates due to forces, whose directions are shown with arrows coming out of actin polymerization and myosin contractility respectively. $r_0$ is the radius of the undeformed sphere.}
\label{fig:mod_cartoon}
\end{figure}

In addition to the equilibrium forces, actin polymerization from APs against the cortex, with a polymerization rate $f_r$, generates a reaction force pushing the membrane along the local outward normal~(Fig.\ref{fig:mod_cartoon}($b$)\,). Thus $f_r \psi$ corresponds to a force imbalance causing deformation of the membrane. 
In addition, the inward growing actin filaments experience active inward pull with a rate $f_p$ due to the contractile activity of actomyosin~\cite{Doherty2008, Medeiros2006} via attached myosin of density $\varphi(\o)$. This force propagates to the membrane via the actin network~(Fig.\ref{fig:mod_cartoon}($b$)\,). We model this using a spread function. In a highly branched network the myosin pull can affect a relatively wider region of the membrane. However, when the filaments are bundled, it transmits to highly localized parts of the membrane~\cite{Svitkina2003}.
It is reasonable to assume that the maximum of such force on the membrane would be located at the same orientational location $\Omega$ as the site of  force generation in the cortex. We use a Gaussian propagator in the spherical polar coordinates  
\beq
G_\a(\Omega, \Omega')=\sum_{lm}Y^*_{lm}(\Omega')Y_{lm}(\Omega)e^{-\a\, l(l+1)},
\label{gaussiansphere}
\eeq
where $\a$ parameterizes the angular spread over which the active contractile force gets distributed. 
Note that in the limit of $\a = 0$, the completeness condition of  spherical harmonics makes the Greens function $G_0 = \d(\Omega-\Omega')$ absolutely localized  in orientation. 
The resultant contraction on the membrane 
is $\int G_\a(\Omega, \Omega')  [- f_p ] \varphi(\Omega')d\Omega'$.  Thus, the total rate of deformation due to activity is given by,
\beq
f_{\rm act} (\Omega)= f_r\psi(\Omega) -\int G_\a(\Omega, \Omega') f_p \varphi(\Omega')d\Omega' .
\label{myosinforce1}
\eeq
\subsection{Coupled dynamics}
The non-conservative dynamics of the membrane can be expressed as 
$\p u / \p t = -\G \left[ \d F/\d u \right] + f_{\rm act} +\eta_u(\Omega, t)$, where
$\eta_u$ is a thermal noise obeying 
$\la \eta_u(\Omega, t) \ra =0$, 
$\la  \eta_u(\Omega, t)   \eta_u(\Omega', t') \ra = 2 \G \kb T \d(\Omega-\Omega') \d(t-t')$. 
In the absence of active forces, these dynamics lead to equilibrium fluctuations. 
The evolution of the displacement field $u(\Omega,t)$ within linear order of $u$, $\psi$ and $\varphi$ fields have the form
\bea
\frac{\partial u}{\partial t} &=& -\G\left[4\k \left(u+\Delta_2 u+\f{1}{4}\Delta_2^2 u\right)
-2\sigma r_0^2 \le( u+\hf\Delta_2u \ri)  \right. \nn\\
&& \left. + K_t r_0^4 u - \k \bar H r_0 \left( \psi +\hf \D_2  \psi\right) \ri] \nn\\
&& -\int d\Omega' G_\a(\Omega,\Omega')f_p \varphi(\Omega')  +f_r \psi + \eta_u(\Omega,t) .
\label{udot}
\eea
We assume that the number of APs on the membrane is conserved. Its dynamics can be expressed as 
$\frac{\partial \psi}{\partial t}=\mu\Delta_2\frac{\delta F}{\delta \psi} + \eta_\psi(\Omega,t) $
where $\mu$ is a transport coefficient describing the angular mobility of APs, and the stochastic noise obeys the relations 
$\la \eta_\psi(\Omega,t)\ra =0$, 
$\la \eta_\psi(\Omega,t) \eta_\psi(\Omega',t') \ra = - 2\mu \kb T \D_2 \d(\Omega-\Omega') \d(t-t') $. 
Performing the functional differentiation and keeping terms up to linear order 
\beq
\frac{\partial \psi}{\partial t}= D \D_2 \psi - \mu \k \bar H r_0 \le(\D_2 u + \hf \D_2^2 u \ri) +  \eta_\psi(\Omega,t),
\label{eomn2}
\eeq
with angular diffusion constant $D= \mu \L r_0^2$. 
The cortical myosin undergoes turnover between a passive detached state to an active cross-linked state in which it can pull the actin filament grown from the APs inward. The variation of myosin density can be described by 
$\frac{\partial \varphi}{\partial t}=-k_{\rm off} \varphi+k_{\rm on} \psi + \eta_\varphi(\o,t),$
where $\eta_\varphi$ denotes thermal noise in the myosin on-off rate~\cite{Shlomovitz2007}. 
In writing this equation, $\psi$ is used as a proxy to the availability of F-actin grown from the APs. The attachment rate $k_{\rm on}$ of myosin to F-actin is diffusion-limited, and is assumed to be constant. 
The detachment rate may increase with local load $f_\ell$  as $k_{\rm off} = k_0 \exp(f_\ell/f_d)$ where $f_d$ sets the scale of detachment force. The elastic load force due to membrane deformation  $f_\ell \sim | u |$. Thus up to linear order, we replace $k_{\rm off} \varphi$ by $k_0 \varphi$, to get 
\beq
\frac{\partial \varphi}{\partial t}=-k_0 \varphi+k_{\rm on} \psi + \eta_\varphi(\o,t). 
\label{eq_m}
\eeq
At steady state $\varphi_0 = (k_{\rm on}/k_0) \psi_0$ with  $\psi_0$ being the average concentration of APs. In the following we consider small deviations $\psi,$ $\varphi$, and $u$  around $\psi_0$, $\varphi_0$ and $u=0$.
\section{Results}
\label{results}
We expand the fields of small deviations $u,\, \psi,\, \varphi$  in the basis of the spherical harmonics $Y_{lm}(\Omega)$ with amplitudes $u_{lm} (t)$, $\psi_{lm} (t)$, and $\varphi_{lm} (t)$~(see Appendix-\ref{appendixC} and \ref{app:dyn}). Within the mean-field approximation, ignoring stochastic noise, the equations (\ref{udot}), (\ref{eomn2}) and (\ref{eq_m}) lead to the dimensionless form, 
\begin{eqnarray}
\f{d u_{lm}}{d \t} &=& -4\tilde{\kappa}\left[\frac{l(l+1)}{2}-1\right]^2u_{lm}-2\tilde{\sigma}\left[\frac{l(l+1)}{2}-1\right] u_{lm} \nn\\
&-& \tilde{K_t}u_{lm} +\tilde{f_r}{\psi}_{lm}-\tilde{\kappa}\tilde{H}\left[\frac{l(l+1)}{2}-1\right] \psi_{lm} \nn\\
&&-\tilde{f_p}e^{-\alpha l(l+1)}{\varphi}_{lm}\nonumber\\
\f{d {\psi}_{lm}}{d \t}&=&-l(l+1) \left\{ {\psi}_{lm}- \tilde{\kappa}\tilde{H}\tilde{\mu} 
\left[ \f{l(l+1)}{2}-1\right]u_{lm} \right\}\nonumber\\
\f{d {\varphi}_{lm}}{d \t} &=& -\tilde{k}_0{\varphi}_{lm} + \tilde{k}_{\rm on}{\psi}_{lm},
\label{dimless_time_evo}
\end{eqnarray}
where $\tau = tD$, $\tilde{\kappa} = \frac{\Gamma\kappa}{D}$, $\tilde{\sigma} = \frac{\sigma\Gamma r_0^2}{D}$, $\tilde{K_t}=\f{K_t r_0^4\G}{D}$, 
$\tilde{f_r}=\frac{f_r}{D}$, $\tilde{f_p} = \frac{f_p}{D}$, $\tilde{H} = \bar{H}r_0$, $\tilde{\mu} = \frac{\mu}{\Gamma}$, $\tilde{k}_0 = \frac{k_0}{D}$ and $\tilde{k}_{on} = \frac{k_{on}}{D}$. 
The volume constraint within quadratic order in $u$ gives the relation $\int u~d\o = -\int u^2~d\o$. In terms of spherical harmonics this leads to $\sqrt{4\pi}~ u_{00} = - \sum_{l,m} u_{l,m}^2$, i.e., $u_{00}$ is a term of quadratic order in $u_{l,m}$. Similarly, the constraint on AP number leads to $\sqrt{4\pi}~ \psi_{00} = \hf\sum_{l,m} [2-l(l+1)] u_{lm}^2 - 2 \sum_{l,m} u_{l,m} \psi_{l,m}$, i.e., $\psi_{00}$ is of quadratic order in $u_{l,m}$ and bilinear order in $u_{l,m}$ and $\psi_{l,m}$. As a result, $u_{00}$ and $\psi_{00}$ are negligible in linear order. We use $u(\o,t) = \sum_{l\geq 1, m} u_{lm}(t) Y_{lm}(\o)$ and $\psi(\o,t) = \sum_{l\geq 1, m} \psi_{lm}(t) Y_{lm}(\o)$. At the steady state $\varphi_{lm} = (\tilde k_{\rm on}/\tilde k_0) \psi_{lm}$ requires $\varphi(\o,t) = \sum_{l\geq 1, m} \varphi_{lm}(t)  Y_{lm}(\o)$.

In the following, we perform a stability analysis of Eq.(\ref{dimless_time_evo}) to determine the dynamical phase behaviors depending on the various control parameters.

\subsection{Stability analysis and dynamical phases} 
\label{subsec:lsa}
We can determine the evolution of the vector $| \Psi \ra_{lm} = (u_{lm}, \psi_{lm}, \varphi_{lm})^T$ analyzing the properties of the stability matrix ${\bf S}^l$ governing the evolution in Eq.(\ref{dimless_time_evo}), $d| \Psi \ra_{lm}/d\t = {\bf S}^l |\Psi \ra_{lm}$. The stability matrix is a function of $l$ alone, as a result, the evolution of $|\Psi \ra_{lm}$ is independent of $m$.
 The eigenvalues can be determined from the characteristic equation
\bea
 \l^3 + a_l \l^2 + b_l \l  + c_l = 0 
\label{eq:char}
\eea 
where $a_l = - {\rm Tr} ({\bf S^l})$, $b_l = \hf \left( { s^l_{ii}} { s^l_{jj}} - {  s^l_{ij}}{  s^l_{ji}}\right)$ with ${ s^l_{ij}}$ denoting elements of the matrix ${ {\bf S}^l}$, and $c_l = - {\rm det}(\bf S^l)$. A  summation over repeated indices is assumed. The detailed expressions for the components of ${\bf S}^l$, and $a_l$, $b_l$, $c_l$ are listed in Appendix-\ref{app_m}.

\begin{figure}[t]
\begin{center}
\includegraphics[width=9cm]{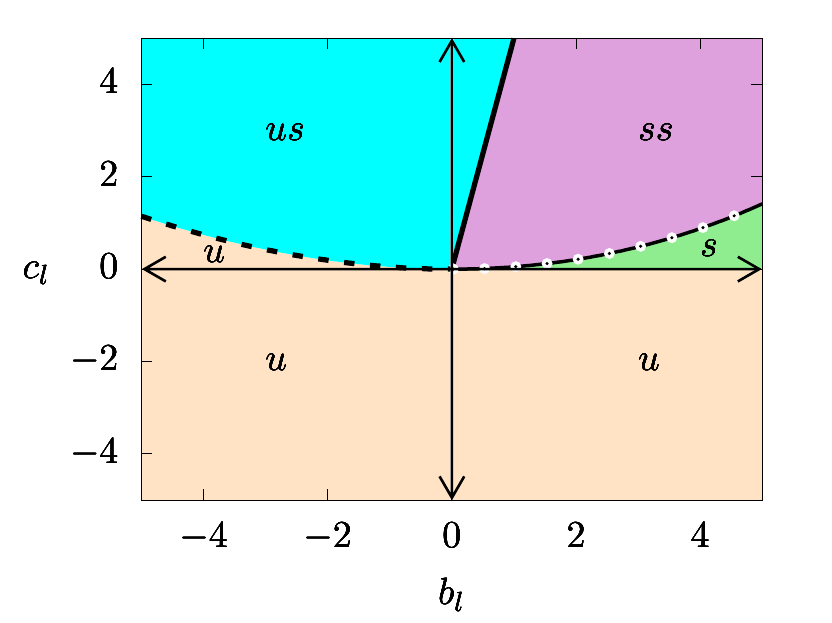}
\end{center}
\caption{The schematic diagram shows different possibilities of phases in the $b_l$-$c_l$ plane, for a fixed $a_l=5$.
The solid line denotes the phase boundary between the stable-spiral ($ss$) and unstable-spiral ($us$) phases. The dash-dotted (dashed) line denotes the phase boundary between the linearly stable $s$ (unstable $u$) and the stable-spiral $ss$ (unstable-spiral $us$)  phase in the first (second) quadrant. The $c_l=0$ line denotes the linearly stable $s$ to unstable $u$ phase boundary in the first quadrant. }
\label{fig:phase_scheme}
\end{figure}
As can be seen from these expressions 
$a_l > 0$ always, while $b_l$ and $c_l$ can change sign. The possible dynamical phases in the $b_l$-$c_l$ plane can be determined in terms of the three roots of the cubic polynomial $p(\l) = \l^3 + a_l \l^2 + b_l \l$ obeying the condition $p(\l) = - c_l$. For $b_l > 0$, one of the roots remains real negative (positive), if $c_l > 0$~($c_l < 0$). The other two roots could be either real negative, or appear as complex conjugate pairs. The  dynamics corresponding to all real negative (or at least one positive) roots is linearly stable (unstable). On the other hand, the dynamics displays stable spiral (unstable spiral) behavior if the real part of the complex conjugate solutions, when they are present, is negative (positive). These conditions can be directly verified by noting that  in terms of the solutions $\l_{1,2,3}$, the coefficients must obey 
 $a_l = - (\l_1 + \l_2 +\l_3)$, $b_l = \l_1 \l_2 + \l_2 \l_3 + \l_3 \l_1$ and $c_l = -\l_1 \l_2 \l_3$. 
  
 The above-mentioned dynamical phases and phase- boundaries  in the $b_l$-$c_l$ plane are illustrated in Fig.\ref{fig:phase_scheme} using a fixed $a_l = 5$. In the first quadrant, $b_l>0$, a phase boundary between the linearly stable ($s$) and linearly unstable ($u$) phases appears at 
 \bea
 c_l=0.
 \eea 
 In the regime of $b_l>0$, another phase boundary between the linearly stable ($s$) to stable spiral ($ss$) phase appears at 
 \bea
 c_l = \left( \f{a_l}{3} + \f{2}{3} \sqrt{a_l^2 - 3 b_l} \right) \left( -\f{a_l}{3} + \f{1}{3} \sqrt{a_l^2 - 3 b_l} \right)^2.
 \eea
 Both these phases are stable in the long time limit. 
 The same equation, however, describes a more interesting phase boundary between the linearly unstable ($u$) and the unstable spiral ($us$) phase, for $b_l<0$. Finally, the boundary between the stable spiral ($ss$) and unstable spiral ($us$) phase is described by 
 \bea
 f_l := c_l - a_l b_l =0.
 \label{eq:ssus} 
 \eea 
 The instabilities determine the shape deformations~\footnote{The diagonalization of ${\bf S}^l$ via the similarity transformation ${\bf D}$ can be used to obtain the time-dependent solution in terms of  the stochastic noise. Writing 
 $| \Phi \ra_{lm} = {\bf D} | \Psi \ra_{lm} $, noise $| \zeta \ra_{lm} = {\bf D} | \eta \ra_{lm} $ and
 ${\bf D} {\bf S}^l {\bf D}^{-1} = \l {\bf 1}$,  Eq.(\ref{dimless_time_evo}) leads to 
 $d | \Phi \ra_{lm}/d\t = \l | \Phi \ra_{lm}+ | \zeta \ra_{lm} $ providing solution 
 $ | \Phi (\t)\ra_{lm} = e^{\l {\bf 1} (\t-\t_0)} | \Phi (\t_0)\ra_{lm}+ \int_{\t_0}^\t e^{\l {\bf 1}(\t - \t')} | \zeta (\t')\ra_{lm} d\t' $. The existence of instabilities quickly starts to dominate.
 }.
 Further details of the determination of these phase boundaries are described in Appendix-\ref{ap:linstab}.
 The growing instabilities within linearized analysis are expected to get saturated at late times  due to the non-linearities in the system, e.g., associated with bending energy and surface energy costs~(Appendix-\ref{appendixA}). As a result they can lead to pattern formation in $u$ phase and limit cycle oscillations in the $us$ phase~\cite{Strogatz2014, cross_greenside_2009}.

In the following, using the generalized phase diagram obtained in the $b_l$-$c_l$ plane, we explore the specific dynamical phase transitions by varying $f_r$, $f_p$, and $\s$, keeping all other parameters fixed at biologically realizable values~(Table-\ref{table_1}). The volume constraint ensures the absence of spherically symmetric expansion or compaction corresponding to the $l=0$ mode. As we show in Appendix-\ref{appendix_l0_l1} the possible deformations due to $l=1$ mode is always stable. The morphological changes involve only the higher $l$- modes.
 
\begin{table}[h]
\caption{In this table we specify the exact values of the dimensionless parameters that have been used in our calculations. Throughout the paper we have kept the values of $\tilde{\kappa}$, $\tilde{k}_{on}$, $\tilde{k}_0$ and $\alpha$ fixed as given here. Values marked with $(A)$ correspond to instabilities involving unstable spirals (e.g., Fig.\ref{fig:phase_dia_sig_pus2}) and those marked with $(B)$  correspond to pattern formation (e.g., Fig.~\ref{fig:phase_dia_sig_pul1}). A discussion on the choice of parameter values and  their full list is provided in the ESI.}
\centering
\begin{tabular}{l l l}
\hline\hline
Bending modulus & $\tilde{\kappa}$& $25$~~\cite{Shlomovitz2007}\\ 
Attachment rate   & $\tilde{k}_{on}$ & $3\times 10^4$~~\cite{Shlomovitz2007,Chen2009a} \\
Bare detachment rate & $\tilde{k}_0$ & $10$~~\cite{Shlomovitz2007}\\
Angular spread of myosin force & $\alpha$ & $0.001$ \\
AP induced curvature & $\tilde{H}$ & $-1^{(A)},~1^{(B)}$~~\cite{Shlomovitz2007}\\
 AP mobility coefficient & $\tilde{\mu}$ & $0.15^{(A)}$, $0.02^{(B)}$\\
 Tether & $\tilde{K_t}$ & $1000^{(A)},~2000^{(B)}$\\
 \hline\hline
\end{tabular}
\label{table_1}
\end{table}%

 \begin{figure*}[t]
 \begin{center}
 \includegraphics[width=5.8cm]{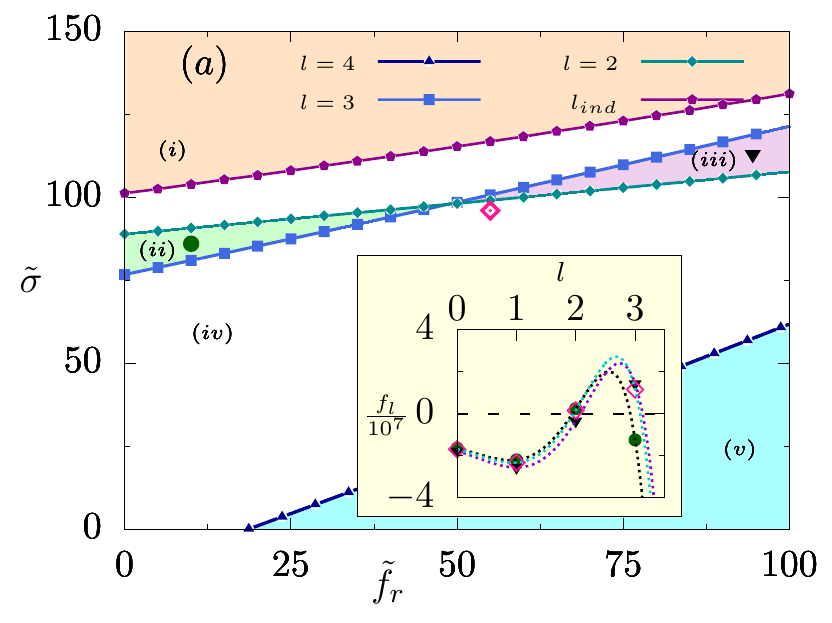}
  \includegraphics[width=5.8cm]{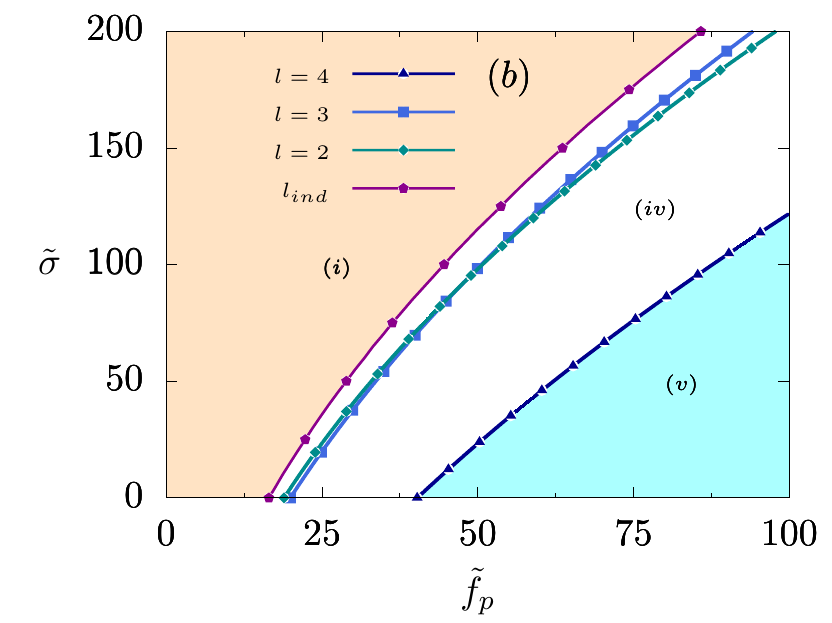}
  \includegraphics[width=5.8cm]{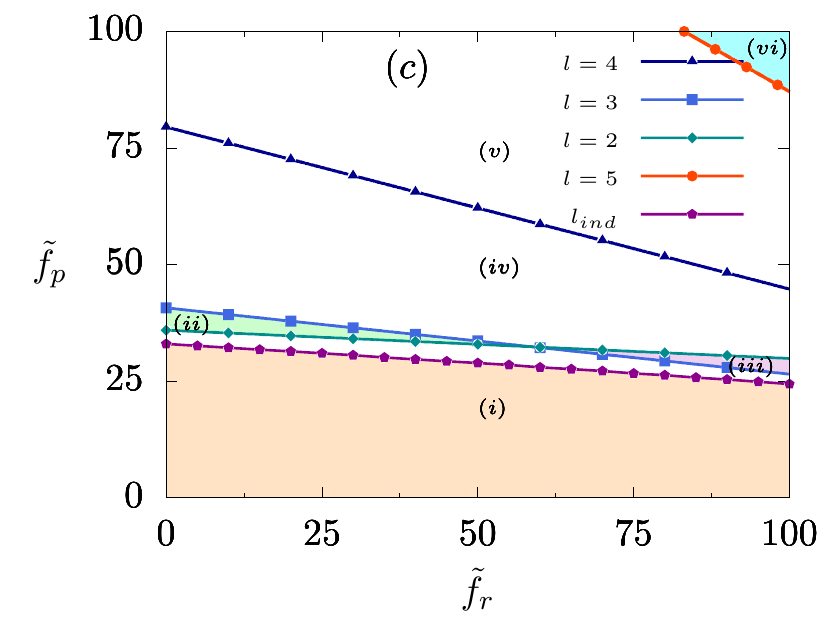}
 \end{center}
 \caption{Phase diagrams with AP-coupling $\tilde{H}=-1$ showing 
 transitions between stable spiral ($ss$) and unstable spiral ($us$) phases, at fixed parameter values $\tilde{\mu} = 0.15$, $\tilde{K_t} = 10^3$. In the three phase diagrams  ($a$) $\tilde f_p=50$, ($b$)~$\tilde f_r=50$ and ($c$)~$\tilde\s=50$ are kept fixed. In figures ($a$) to ($c$) different regions and lines denote the following. The analytical estimate for the onset of $us$ phase independent of $l$-modes, treating $l$ to be a real number, is shown by the line $l_{ind}$. The whole region ($i$) shaded orange corresponds to $ss$ phase for all $l$. Regions ($ii$) (shaded green) and ($iii$) (shaded violet) show $us$ behavior separately for $l=2$ and $3$, respectively. In region ($iv$), both $l=2$ and $l=3$ modes display $us$. Region ($v$) (shaded cyan) corresponds to $us$ behavior for $l=2,\,3,\,4$ modes. 
 Moreover, the region $(vi)$ in figure ($c$) displays $us$ for all of $l=2,\,3,\,4,\,5$ modes. 
 The color shades of different regions are the same in figures ($a$), ($b$), and ($c$). 
 The phase boundary between the localized  and running pulsations lies in between the points denoted by red $\Diamond$ corresponding to the running pulsation and green $\Circle$ (black $\triangledown$) associated with the localized pulsation dominated by $l=2$ ($l=3$) mode. 
 The region of localized pulsation is above this line in figures ($a$) and ($b$), while it remains  below this line in figure ($c$).
  }
 \label{fig:phase_dia_sig_pus2}
 \end{figure*}

\begin{figure}[!t]
\begin{center}
\includegraphics[width=8.6cm]{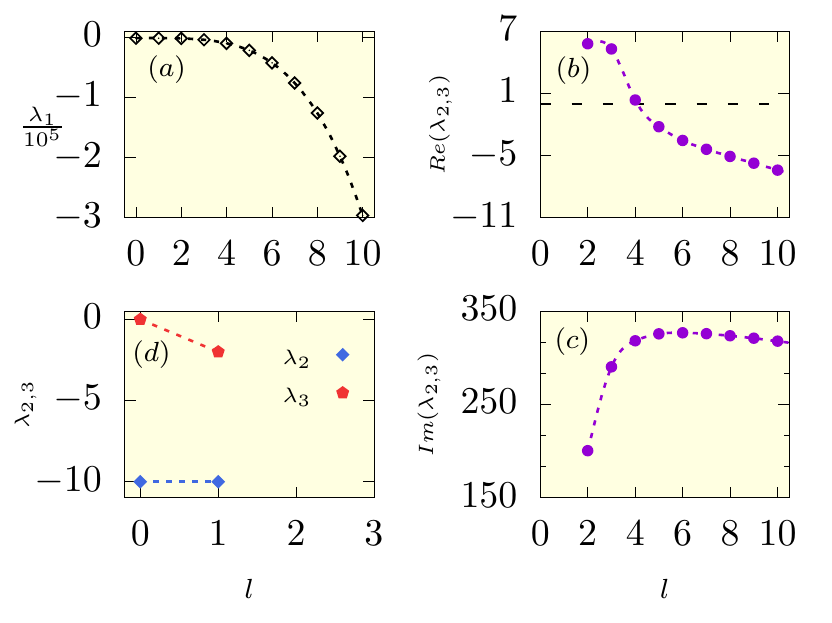}
\end{center}
\caption{Plots showing the variation of eigen-values $\lambda_{1,2,3}$. The parameter values are $\tilde{H}=-1$, $\tilde{\mu} = 0.15$, $\tilde{K_t} = 10^3$, $\tilde f_p=50$, $\tilde{\sigma} = 30$ and $\tilde{f_r} = 70$~(region ($v$) of Fig.~\ref{fig:phase_dia_sig_pus2}($a$)\,). 
$\l_1$ remains real negative for all $l$~($a$). $\l_{2,3}$ are real negative at $l=0,\,1$~($d$), and shows complex conjugate values at higher $l$. The real and imaginary parts  of $\l_{2,3}$ for $l>1$ are shown in ($b$) and ($c$). The $l=2$ mode is maximally unstable with largest positive value of $Re(\l_{2,3})$ as shown in ($b$). 
$Im(\lambda_{2,3})$ identify the frequency of oscillations of different $l$ modes~($c$). 
}
\label{fig:eigen_var_l_70_30_1}
\end{figure}
\subsection{Pulsatory deformations}  
We first consider a negative coupling $\tilde{H}=-1$ between AP and membrane curvature, such that APs prefer and induce local {\em hills}  on the membrane. 
In Fig.~\ref{fig:phase_dia_sig_pus2}($a$)--($c$) we show phase diagrams identifying various $us$ behaviors in the small $l$ limit. The obtained phase behaviors are due to the  competition between active forces $\tilde f_p$, $\tilde f_r$ and passive stabilizing factor $\tilde \s$.
The parameters here maintain the relations $b_l > 0$ and $c_l>0$, with the transition lines denoted by $f_l=0$, where  $f_l:=c_l-a_l b_l$. The system remains in the $ss$-phase for $f_l<0$, and transits to $us$ phase for $f_l>0$.  The function $f_l$ varies non-monotonically with a maximum at $l = l_{\rm max}$ (Fig.~\ref{fig:fl} in Appendix-\ref{app_fl}) displaying the presence of a band of $l$ corresponding to the $us$ phase, such that the system is stable ($ss$) for all other $l$-values. 
The deformations displayed in the $us$-phase thus depend on which of the $l$-modes are excited in the system.  
Setting $f_{l_{\rm max}}=0$ gives an estimate of the phase boundary, above (below) which all $l$-modes are stable ($ss$) in figure ($a$) and ($b$) (figure ($c$)\,). This is shown by the beaded line denoted by $l_{ind}$ in Fig.\ref{fig:phase_dia_sig_pus2}($a$)--($c$).  However, $l$ is  an integer and not a real number, as a result making this estimate only approximate.  The regions shaded orange in Fig.\ref{fig:phase_dia_sig_pus2} denote the $ss$ phase.
Due to the presence of unstable bands, the different regions in Fig.\ref{fig:phase_dia_sig_pus2} correspond to $us$- phase associated with $l=2$~($ii$), $l=3$~($iii$), $l=2,3$~($iv$), $l=2,3,4$~($v$), and $l=2,3,4,5$~($vi$) modes.

 The existence of an unstable $l$ band gets reflected in the dependence of eigenvalues $\l_{1,2,3}(l)$.
 We show this in Fig.\ref{fig:eigen_var_l_70_30_1} using $\tilde \s=30$ and $\tilde f_r=70$, keeping all other parameter values the same as in Fig.\ref{fig:phase_dia_sig_pus2}($a$). 
Fig.\ref{fig:eigen_var_l_70_30_1}($a$) shows $\l_1 < 0$ for all $l$. On the other hand $\l_{2,3}$ are real negative for $l=0,1$ (Fig.\ref{fig:eigen_var_l_70_30_1}($d$)\,), while they show complex conjugate values for $l>1$. Fig.\ref{fig:eigen_var_l_70_30_1}($b$) shows that the real parts of $\l_{2,3}>0$ only for $l=2,\,3,$ and 4, supporting a narrow band of instability in $us$ phase. For $l>4$ they become negative, and the system gets into $ss$ phase. The imaginary part of $\l_{2,3}$ shown in Fig.\ref{fig:eigen_var_l_70_30_1}($c$) determines the frequency of oscillations. 
 
 \begin{figure}[!t]
 \begin{center}
 \includegraphics[width=8.6cm]{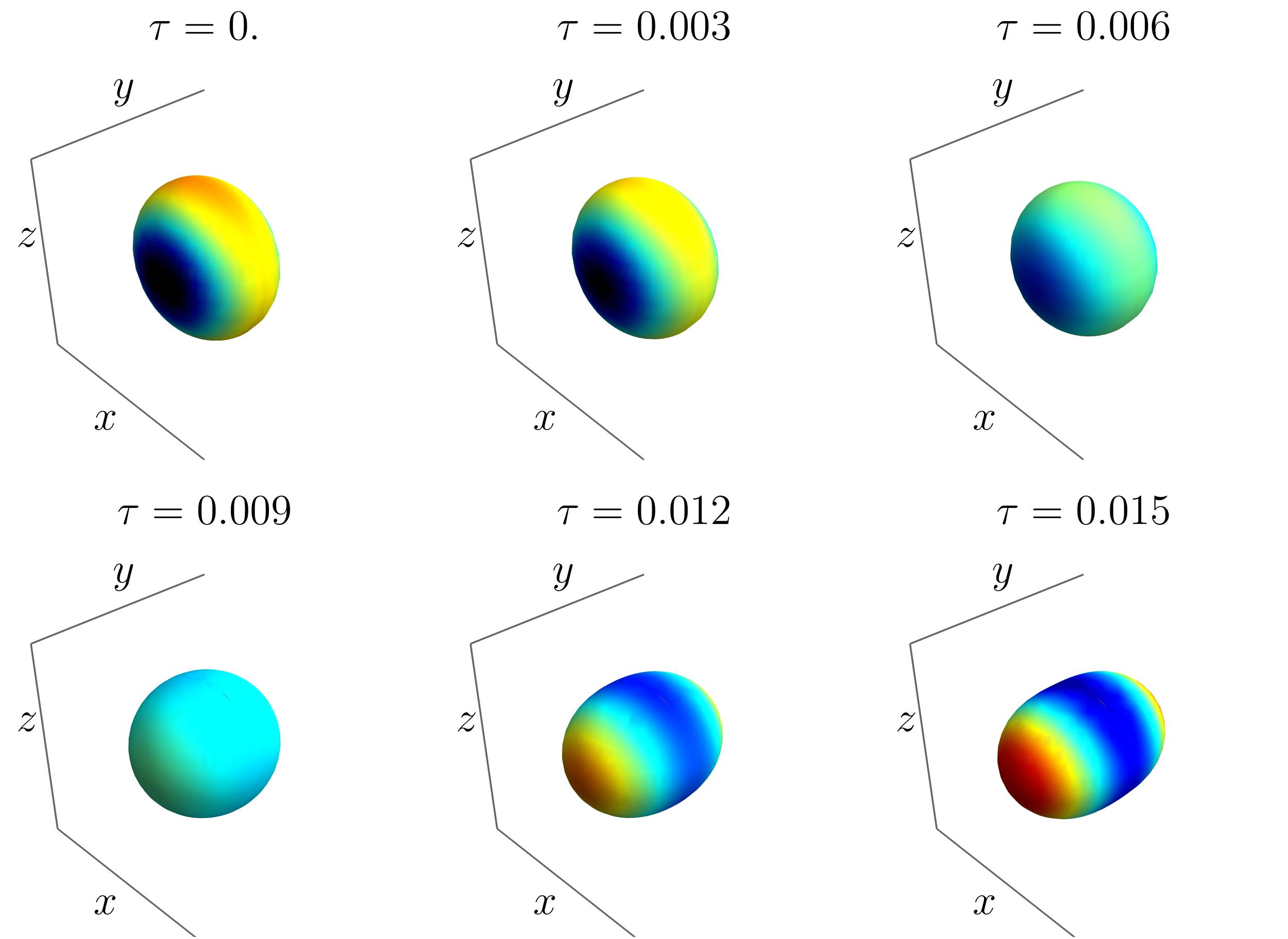}
 \includegraphics[width=6cm]{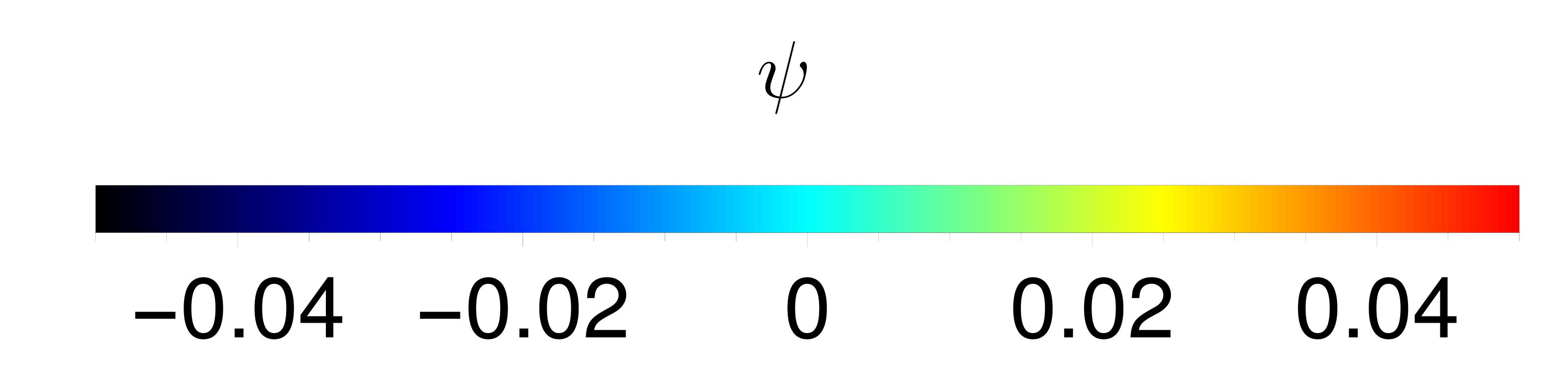}
  \end{center}
 \caption{Plots showing a half cycle of a localized pulsation governed by $l=2$ mode on a spherical membrane when the APs prefer hills $\tilde H=-1$ in the presence of a strong myosin contractility. The color code on the deforming spherical shapes denotes  the local AP concentration $\psi$. The parameter values are $\tilde{H}=-1$, $\tilde{\mu} = 0.15$, $\tilde{K_t} = 10^3$, $\tilde f_p=50$,  $\tilde{\sigma} = 86$ and $\tilde{f_r} = 10.0$.
 The time for each snapshot is denoted by $\t$ values shown in the
 figure. The snapshots here correspond to the movie local\_puls.avi in the ESI.}
 \label{fig:time_evo_os_l2}
 \end{figure}

Fig.~\ref{fig:phase_dia_sig_pus2} predicts the onset of $us$- phase corresponding to different $l$-modes. Moreover, the detailed numerical analysis show two different natures of the $us$ phase, characterized by spatially localized pulsation and running  pulsation. 
The boundary between these two kinds of pulsations lies in between open $\Diamond$ symbols and filled $\Circle$ (filled $\triangledown$) for $l=2$ ($l=3$) mode. 
We present their detailed characterization, in the following.

 \begin{figure}[!t]
 \begin{center}
 \includegraphics[width=8cm]{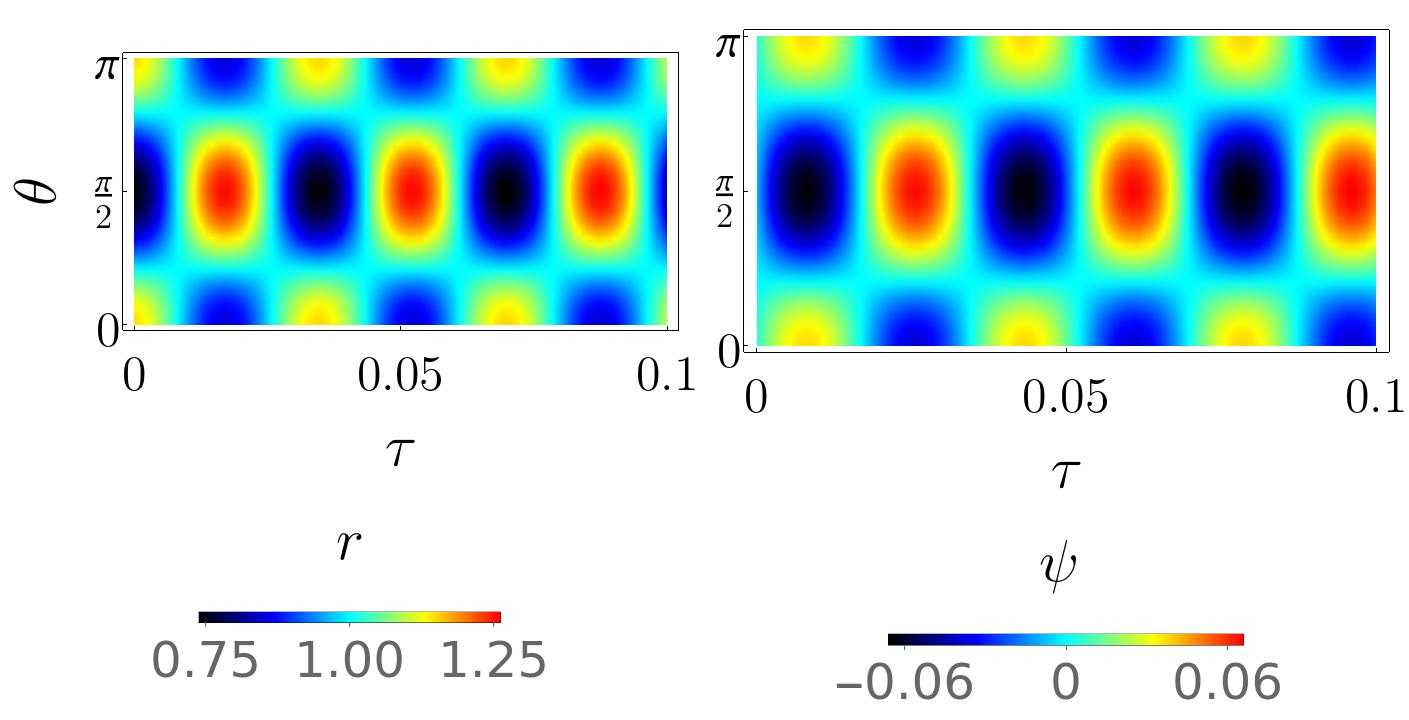}
 \end{center}
 \caption{Localized pulsation: Kymographs depicting localized pulsations in $r=r_0(1+u)$ and $\psi$ along the polar angle $\theta$ at a fixed azimuthal angle $\phi = \frac{\pi}{2}$, at the same parameter values as in Fig.\ref{fig:time_evo_os_l2}. The system excites only the $l = 2$ mode to $us$ phase.  The parameter values are $\tilde{H}=-1$, $\tilde{\mu} = 0.15$, $\tilde{K_t} = 10^3$, $\tilde f_p=50$,  $\tilde{\sigma} = 86$ and $\tilde{f_r} = 10.0$.
 }
 \label{fig:kymo_comp_l2}
 \end{figure}
\subsubsection{Localized pulsations}
In Fig.~\ref{fig:time_evo_os_l2} we show membrane deformations along with AP localization corresponding to region ($ii$) of Fig.~\ref{fig:phase_dia_sig_pus2}($a$), using particularly $\tilde{f_r} = 10.0$, $\tilde{\sigma} = 86$. The fact that these deformations correspond to localized pulsation is easier to see in the kymographs of $u$ and $\psi$  in Fig.(\ref{fig:kymo_comp_l2})
plotted at a fixed 
$\phi=\pi/2$.  
The localized pulsations in $u$ and $\psi$ appear with a small time-lag, as it takes time for APs to accumulate or disperse and to deform the membrane. 
Time evolution of $l=3$ mode and corresponding kymograph corresponding to $\tilde{f_r} = 94.5$, and $\tilde{\sigma} = 113$ of Fig.\ref{fig:phase_dia_sig_pus2}($a$), are shown in Fig.s~2 and 3 of ESI.

The physical mechanism behind the localized pulsation can be ascribed to the following competition. 
The membrane-AP coupling $\tilde H <0$ and the actin polymerization support formation of local outward deformation, while the myosin contractility propagating to membrane via the actin network pulls the membrane inward.   
It is worthwhile to note that the outward membrane deformation due to AP accumulation is controlled by the membrane-AP coupling and the bending rigidity of the membrane. The small $l$-modes (large wavelength) of membrane deformation takes longer to change. 
The outward deformations of the membrane accumulates more APs, deforming the membrane further. The time-scale for such accumulation is controlled by the membrane-AP coupling, the bending rigidity, and the mobility of the APs. The outward push due to the actin polymerization reinforces such hill formation. 
On the other hand, the actomyosin contractility mediated by the spread function pulls the membrane inward.  
As a result, the hill turns into a valley, and the APs start to move out to neighboring locations  with a time-lag controlled by the AP-mobility, the membrane-AP coupling and the membrane bending rigidity. This allows the membrane to relax locally towards the spherical shape, as a new hill starts to appear in the neighborhood.
The cycle repeats to sustain oscillations. The specific $l$-mode that is excited controls the shape of the deformation associated with the localized pulsations.  
In regions ($ii$) and ($iii$) in Fig.~\ref{fig:phase_dia_sig_pus2} only one $l$-mode excites $us$ behavior controlling the localized pulsations of the sphere. The phase space regions near them are dominated by these $l$-modes.
In the presence of localized pulsations, the dynamics on closed membrane differs qualitatively from the open flat membranes that cannot support such standing waves but only show traveling waves of membrane protrusions~\cite{Shlomovitz2007,Gov2018}.

\subsubsection{Running pulsations} 
\begin{figure}[!t]
 \begin{center}
 \includegraphics[width=3.5in]{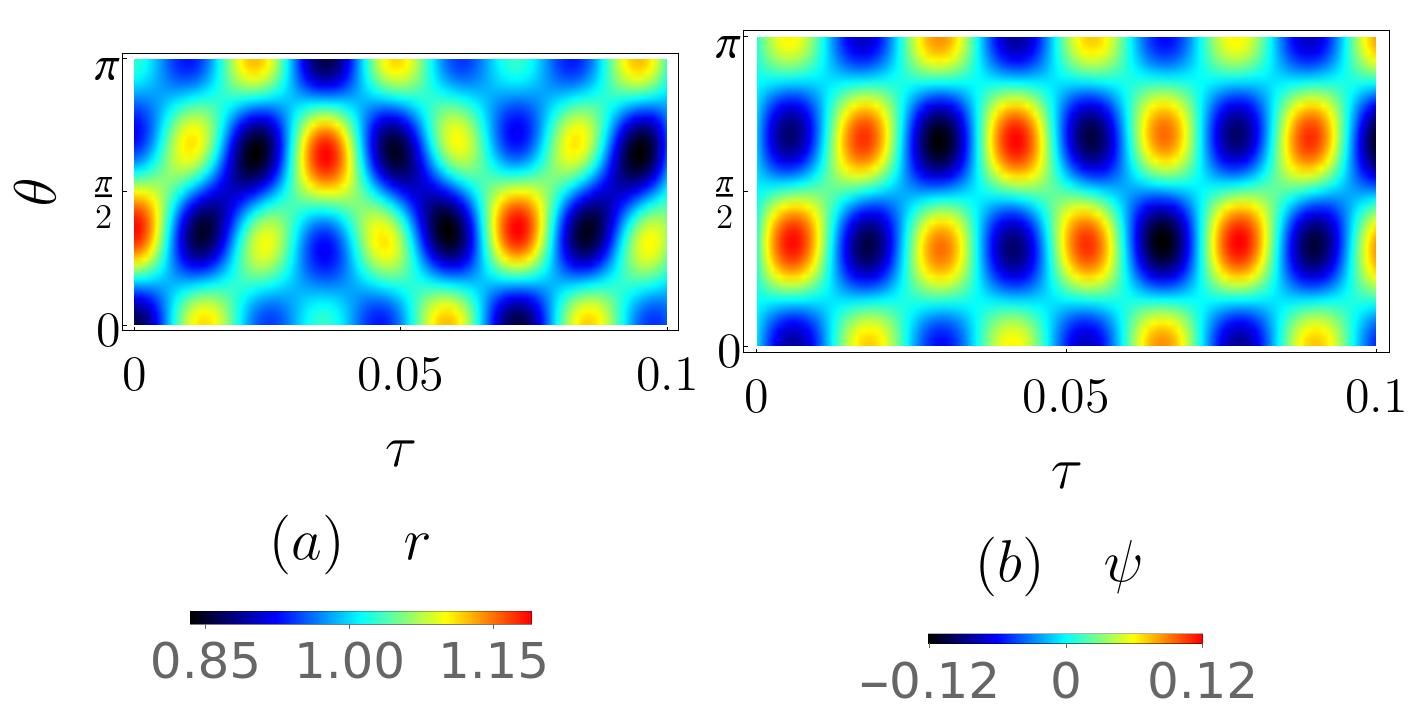}
  \end{center}
 \caption{Running pulsation: Kymographs depicting the time evolution of membrane deformation $r$ and change in AP density $\psi$ along the polar angle at a fixed $\phi = \frac{\pi}{2}$ 
 when a running pulsation is generated due to superposition of $l = 2,~3$ modes on the spherical membrane.  The parameter values are $\tilde{H}=-1$, $\tilde{\mu} = 0.15$, $\tilde{K_t} = 10^3$, $\tilde f_p=50$,  $\tilde{\sigma} = 60$ and $\tilde{f_r} = 75$. 
 The connected nature of deformations and their slope, e.g., the slopes of the troughs in ($a$) capture the traveling wave nature. We show the deformations on the sphere in Fig.~\ref{fig:curr_snap} of Appendix-\ref{app_shapes} and in the movie run\_puls.avi in the ESI.}
 \label{fig:kymo_comp_trav}
 \end{figure}

A remarkable modification of the pulsatory dynamics appears when two modes of similar amplitudes and nearby frequencies are excited together on the closed membrane. For example, we consider the phase-point $\tilde{f_r} = 75$ and $\tilde{\sigma} = 60$ of Fig.~\ref{fig:phase_dia_sig_pus2}($a$\,) where both $l=2$ and $l=3$ modes are excited together in the $us$-phase.  As a result, a running pulsation appears, which can be seen  clearly in the kymograph Fig.~\ref{fig:kymo_comp_trav} to run from pole to pole in a to-and-fro motion. The evolution of the deformations on the sphere is shown in Fig.~\ref{fig:curr_snap} of Appendix-\ref{app_shapes} and movie run\_puls.avi in the ESI. As can be seen clearly from the movie, the pole to pole deformations running on the sphere lead to its forward and backward somersaults with time. 

 The physical mechanism leading to running pulsation is equivalent to that of localized pulsation, in the feedback control and phase lag. The main difference is that here two $l$- modes of similar amplitudes and {\em nearby} frequencies are excited together, and they superpose. As a result, the hills and valleys are formed in a connected manner, in both space and time, allowing the deformations to continuously flow over the membrane surface giving rise to a traveling wave. The small difference in their wavelength ($q^{-1}  \approx r_0/l$) and frequency ($\omega = Im(\l_{2,3})$) maintains a traveling wave packet, whose velocity in the flat membrane limit is given by $d \omega/dq$.  Lateral waves on cell membranes of various cell types, including mouse embryonic fibroblasts, T-cells, and wing disk cells of the fruit fly,  were observed in experiments~\cite{Dobereiner2006}.  
 Localized and running pulsation behaviors have recently been reported for the artificial system of giant unilamellar vesicles encapsulating Min protein system~\cite{Litschel2018a}.

 \begin{figure}[!t]
 \begin{center}
 \includegraphics[width=8cm]{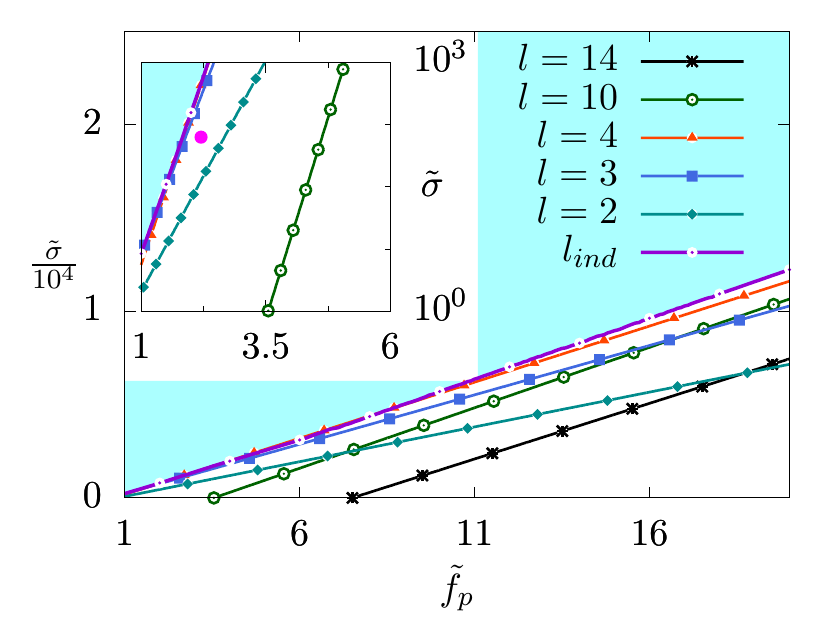}
 \end{center}
 \caption{Phase diagram for the spherical membrane, with valley preferring APs $\tilde{H}>0$ in $\tilde{\sigma}-\tilde{f_p}$ plane where dynamical transition form $s$ to $u$ phase takes place. As the actomyosin pull gets coupled with inward curvature producing APs, it brings about linearly unstable phase in the system through an uncontrolled positive feedback. The parameter values are $\tilde{H} = 1$, $\tilde{\mu} = 0.02$, $\tilde{K_t} = 2000$, 
 $\tilde{f}_r = 10$. The  region with cyan shade above the line $l_{ind}$ is linearly stable for all $l$ modes. Inset: magnification of the phase diagram near small $\tilde f_p$. The dynamics at the point denoted by filled magenta $\Circle$ is illustrated in detail in Fig.s~\ref{fig:time_evo_unstab_l4}, \ref{fig:kymo_unstab_l4}.}
 \label{fig:phase_dia_sig_pul1}
 \end{figure}

\subsection{Pattern formation}
Apart from pulsation, Fig.~\ref{fig:phase_scheme} allows the formation of steady patterns, characterized by linear instability. Here we consider a positive membrane- AP coupling $\tilde{H}=1$, such that APs prefer and induce local {\em valleys} on the spherical membrane. Other parameters are fixed at $\tilde{\mu} = 0.02$, $\tilde{K_t} = 2000$
and $\tilde{f}_r = 10$~(see Table-\ref{table_1}). In Fig.~\ref{fig:phase_dia_sig_pul1}, we explore deformations as a function of surface tension $\tilde \s$ and the rate of contractile pull $\tilde f_p$. Within the parameter-range explored in this figure $b_l>0$. The transition from stable $s$-phase ($c_l>0$) to linearly unstable $u$-phase ($c_l<0$) is determined by the condition $c_l=0$. Here $c_l$ varies non-monotonically with $l$ showing a minimum at $l=l_{\rm min}$. 
If the value at minimum $c_{l_{\rm min}} < 0$, this mode and possibly a band of modes around it obeying $c_l<0$ can show  linear instability towards pattern formation. The resultant deformations of the sphere are controlled by this unstable band. In addition, the condition $c_{l_{\rm min}} = 0$ provides a phase boundary independent of $l$, indicated by the dash-dotted line denoted by $l_{ind}$ in Fig.~\ref{fig:phase_dia_sig_pul1}, such that the region above this line corresponding to larger surface tension $\tilde \s$ is stable for all $l$-modes. Instability appears only below this line. The lines for a constant $l$ in Fig.~\ref{fig:phase_dia_sig_pul1} denote the phase boundaries for particular $l$-modes, with regions above (below) such lines denoting stable (unstable) phase corresponding to the mode. The crossing of different lines shows how instability for $l$-modes shifts from one to another. 

\begin{figure}[!t]
 \begin{center}
 \includegraphics[width=8.6cm]{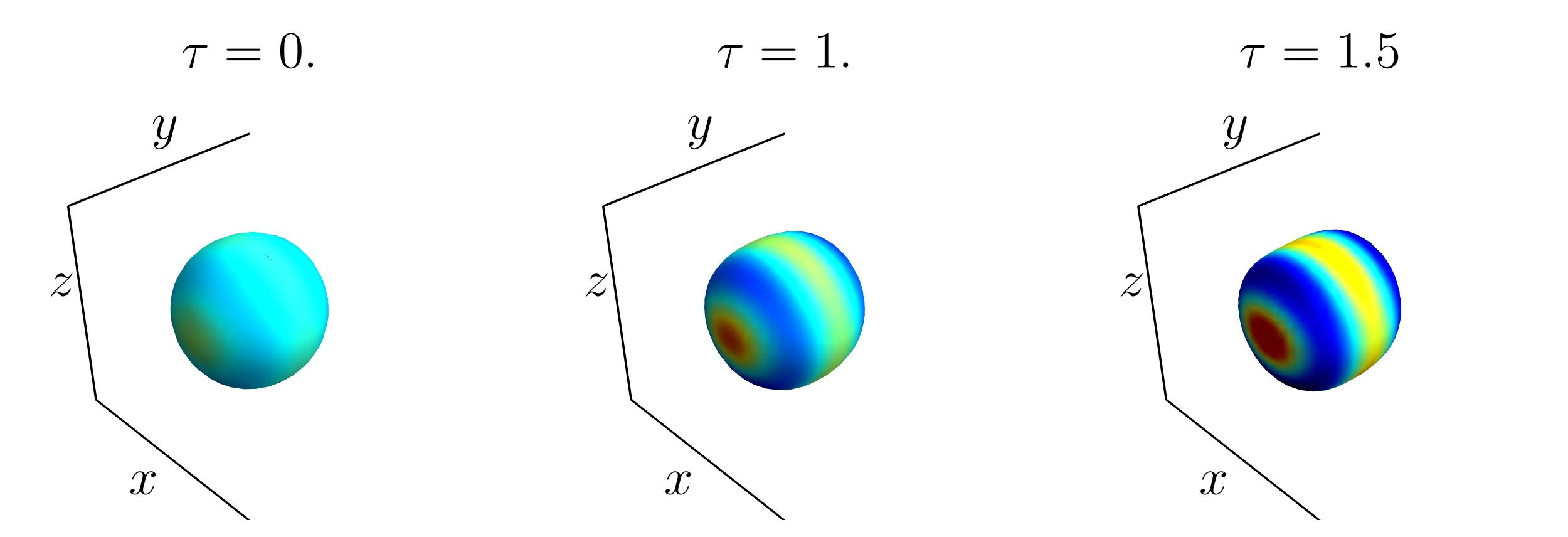}
 \includegraphics[width=6cm]{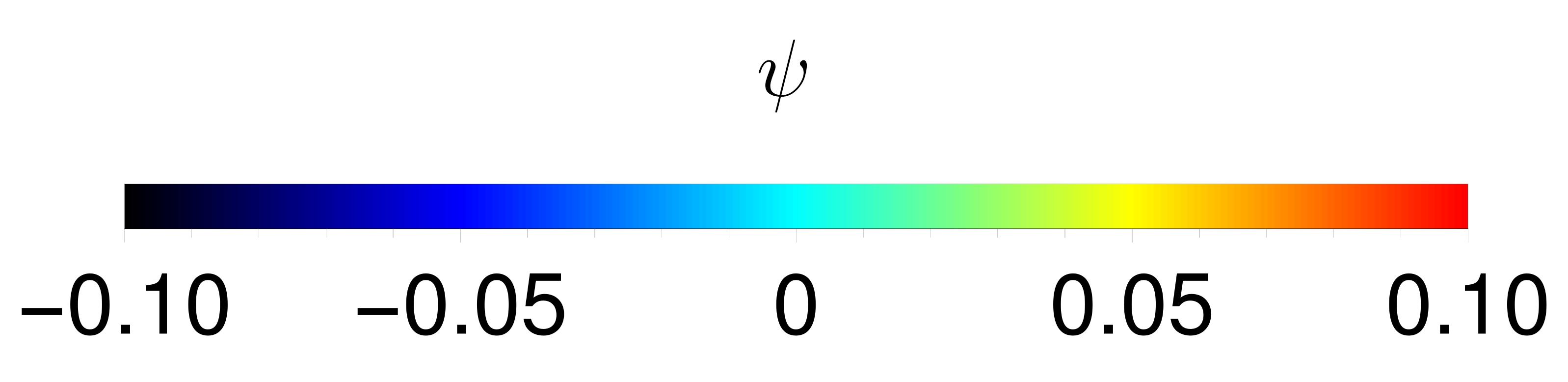} 
 \end{center}
 \caption{Plots showing pattern formation on a spherical membrane due to linear instability of $l=3$ and $4$ modes appearing together, with $\tilde H=1$ so that the APs prefer local valleys. 
The color code on the deforming spherical shapes denotes  the local AP concentration $\psi$.
The parameter values are $\tilde{H} = 1$, $\tilde{\mu} = 0.02$, $\tilde{K_t} = 2000$, $\tilde{f}_r = 10$, $\tilde{\sigma} = 700$, $\tilde{f_p} = 2.2$ corresponding to the point denoted by the filled magenta $\Circle$ in the inset of Fig.~\ref{fig:phase_dia_sig_pul1}.
 The snapshots here correspond to the movie pattern.avi in the ESI. 
}
 \label{fig:time_evo_unstab_l4}
 \end{figure}

\begin{figure}[h]
 \begin{center}
  \includegraphics[width=4.25cm]{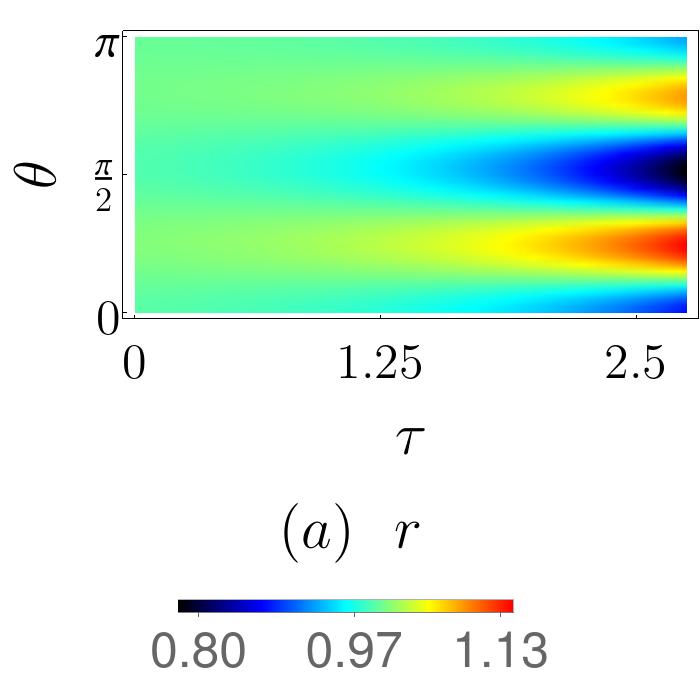}
 \includegraphics[width=4cm]{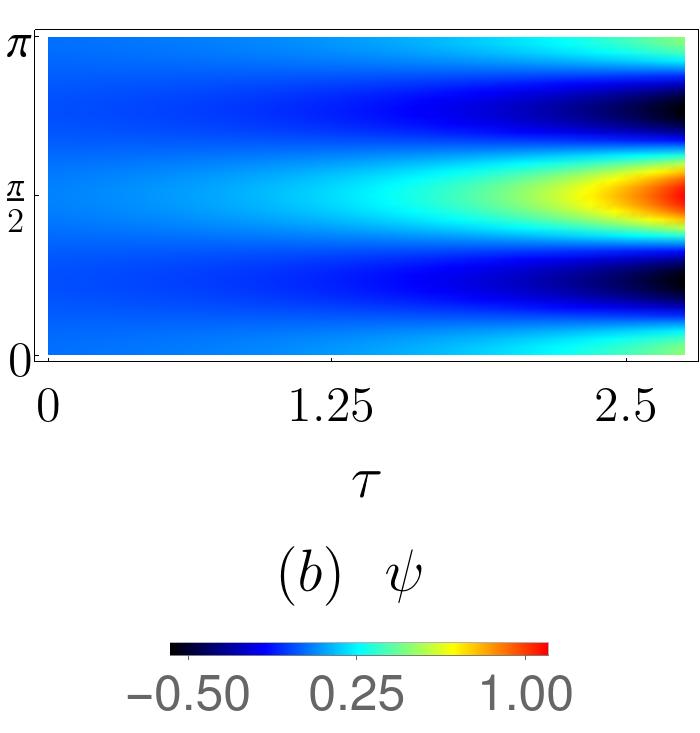}
 \end{center}
 \caption{Instability: Kymographs depicting the onset of instability in the time evolution of membrane deformation $r$ and local AP concentration $\psi$ along the polar angle $\theta$ values at a fixed $\phi = \frac{\pi}{2}$ corresponding to parameter values $\tilde{H} = 1$, $\tilde{\mu} = 0.02$, $\tilde{K_t} = 2000$, $\tilde{f}_r = 10$, $\tilde{\sigma} = 700$, $\tilde{f_p} = 2.2$ as in Fig.~\ref{fig:time_evo_unstab_l4}.
 }
 \label{fig:kymo_unstab_l4}
 \end{figure}

The APs prefer to accumulate at valleys. The F-actins grown from the APs are pulled inward more strongly at larger $\tilde f_p$. This bends the membrane further inward recruiting more APs, a positive feedback mechanism that induces linear instability. This  mechanism of pattern formation is reminiscent of the {\em pump- bump- clump} mechanism described in Ref.~\citenum{Ramaswamy2000}. The initial progression of deformation at $\tilde \s=700$, $\tilde f_p=2.2$ is shown in Fig.~\ref{fig:time_evo_unstab_l4}. At this parameter value, the modes $l=3,\,4$ dominate and determine the observed deformations.  Similar deformations, more so for $l=2$~(Fig.7 in ESI), are observed during cytokinesis in cell division~\cite{Palani2017}. The onset of instability can be seen from the kymograph Fig.~\ref{fig:kymo_unstab_l4}. The high AP density $\psi$ region is associated with inward contraction characterized by small $r$, and vice-versa.   
As expected, deformations at smaller $\tilde \s$ are easier to set in, and instability appears at smaller values of pulling rate $\tilde f_p$. 
 In Fig.~1 of ESI, we show a phase diagram due to the competition between $\tilde \s$ and $\tilde f_r$ for a fixed $\tilde f_p$, displaying transitions between $s$- and $u$-phases. 

\section{Discussion}
\label{discussion}

In summary, we have discussed active deformations of a spherical membrane adhered to a cell cortex. The membrane is described by its surface tension, bending rigidity, and volume constraint -- the shapes of membrane-associated APs couple to the local membrane curvature. We particularly analyzed the impact of active forces due to actin polymerization  and  the contractile actomyosin cortex on the membrane deformation. The large-scale deformation of the spherical membrane and the dynamic pattern formation of AP concentration is studied using linear stability analysis and numerical calculations. 

The generic phase diagram shows possibilities of a stable, unstable, 
stable spiral, and unstable spiral phases. 
A closer investigation within a biologically accessible parameter regime revealed localized and traveling pulsation in the unstable spiral phase and pattern formation due to linear instability. 
The pulsatory patterns are maintained by a negative feedback mechanism between membrane deformation and AP accumulation. We find localized pulsation in the spherical membrane over a broad region of phase space, unlike flat membranes. 
Further, the spherical membrane shows traveling pulsations running between the poles in a to-and-fro motion at parameter values that excite more than one spherical harmonics of similar amplitudes and nearby frequencies to superpose.  
A positive feedback mechanism between membrane deformation and AP accumulation, on the other hand, leads to linear instability. However, the amplitudes of growing instabilities found within the linear stability analysis can get saturated by the stabilizing non-linearities arising from the surface and bending energy costs~\cite{Strogatz2014, cross_greenside_2009}.  
 
Using an expansion in the basis of spherical harmonics, we focussed on the small $l$-modes corresponding to large scale deformations. Some insight into the relative role of different terms governing the coupled dynamics in Eq.(\ref{dimless_time_evo}) can be gained by assuming a fast relaxation for APs and myosin concentrations. Setting $d\psi_{lm}/d\t=0$ we get $\psi_{lm}=\tilde{\kappa}\tilde{H}\tilde{\mu}\left[ \f{l(l+1)}{2}-1\right]u_{lm} $. Further, $d\varphi_{lm}/d\t=0$ gives $\varphi_{lm} = ({\tilde k}_{\rm on}/{\tilde k}_0) \psi_{lm}$. Using these relations  in the first equation of Eq.(\ref{dimless_time_evo}) one can express the {\em slow} evolution of the membrane as
\begin{align}
&\f{d u_{lm}}{d \t} =  -(4\tilde{\kappa} +\tilde\k^2 \tilde H^2\tilde \mu)\left[\frac{l(l+1)}{2}-1\right]^2u_{lm} -\tilde{K_t} u_{lm}\nn\\
&-\left[ 2\tilde{\sigma} +\tilde \k \tilde H \tilde \mu \left({\tilde f}_p e^{-\a l(l+1)} \f{{\tilde k}_{\rm on}}{{\tilde k}_0}  - \tilde f_r \right) \right]\left[\frac{l(l+1)}{2}-1\right] u_{lm}. \nn
\end{align}
The effective bending modulus changes to $\tilde\k+\f{1}{4}\tilde\k^2 \tilde H^2\tilde \mu$ due to the membrane-AP coupling $\tilde H$.
The active forces lead to an effective surface tension $\tilde \s_{\rm eff}=\tilde \s + \tilde \s^{(a)}$, where the active tension is
\bea
\tilde \s^{(a)}=\f{\tilde \k \tilde H \tilde \mu}{2}\left({\tilde f}_p e^{-\a l(l+1)} \f{{\tilde k}_{\rm on}}{{\tilde k}_0}  - \tilde f_r \right).
\label{eq_sa}
\eea
For $\tilde H>0$, $\tilde \s^{(a)}$ increases with the non-local contractile activity $\tilde f_p$, and decreases with the actin  polymerization term $\tilde f_r$. Their role reverses with changing sign of the coupling $\tilde H$. Such mechanisms can be used by the cell to regulate the membrane tension. 

While our analysis incorporating the coupled dynamics of membrane shape, membrane-bound proteins, and active force generators has implications for living cells, our predictions are amenable to direct experimental verification in synthetic cell-like constructs. 
For example, the different aspects of unstable spiral phase can be tested using an artificial cell. 
Depending on the known physical parameters one can obtain quantitative estimates from our model. The viscosity of cytoplasmic extract is $\eta\sim 10\,$mPa-s~\cite{Valentine2005}. 
The F-actin polymerization rate depends on actin concentration. For concentrations up to 4$\mu$M, the F-actin polymerization rate can vary from 2-20 subunits/s~\cite{Kuhn2005}.  Using the subunit size $\sim 2.76$\,nm, this  translates into a growth velocity between $v_g=\,$5.52 nm/s and 82.8 nm/s . 
For a cell of radius $r_0=10\,\mu$m this leads to $f_r=v_g/r_0\approx 5\times 10^{-4} -8\times 10^{-3}$\,s$^{-1}$, which is equivalent to the dimensionless parameter range $0.05 \leq \tilde f_r \leq 0.8$.  
The velocity of F-actin moving on myosin depends on ATP concentration, and the particular type of motor protein within the myosin family.  The skeletal F-actin moving on muscle myosin  can have velocities $v_p$ up to $5\,\mu$m/s~\cite{Kron1986} at large ATP concentrations of $\sim 1\,$mM. 
On the other hand, Myosin XI can have velocities up to $7\,\mu$m/s~\cite{Tominaga2003}, which is equivalent to the dimensionless parameter $\tilde f_p= v_p/r_0 D=70$. 
The membrane-cortex adhesion depends on the ligand and receptor molecules mediating it and their concentrations~\cite{Alert2015}. We considered a weak tethering strength $K_t = 0.001\,$pN/$\mu$m$^3$ describing the adhesion. 
Reduction of the membrane tension by deflating it using hyper-osmotic shock~\cite{Simon2019}, or incorporating cholesterol~\cite{Biswas2019} can assist in its easy deformation to build instabilities.
The phase diagram in Fig.~\ref{fig:phase_dia_sig_pus2}($c$) at $\tilde \s=50$ corresponds to a surface tension $\s=0.005$\,pN/$\mu$m.
At $\tilde f_r=0.1$, one expects to find localized pulsations at $\tilde f_p<45$ and running wave for $\tilde f_p>49$. The predicted oscillations of an artificial cell of radius $r_0=10\,\mu$m will have an amplitude $\sim 1\,\mu$m and a period $\sim 5\,$s. 
At the above-mentioned parameter range, the active tension $\tilde \s^{(a)}
\sim 10^5$~(using Eq.(\ref{eq_sa}) with $\a=0$) shows a $\sim 10^3$ times increase with respect to $\tilde \s$, and corresponds to $\s+\s^{(a)} \approx 10\,{\rm pN/}\mu{\rm m}$, a value within the range of measured effective surface tensions in living cells~\cite{Salbreux2012}.

\appendix
\section{Membrane energetics}
\label{appendix_theory}
Assume a flat membrane in the presence of APs that modifies the local intrinsic curvature and bending stiffness. 
The  APs with an area fraction $\psi$ induces a spontaneous curvature $\psi\, \bar c$, and changes bending rigidity $\k$ to $\bar\k=\k(1-\psi)+\k' \psi$, such that $\bar \k =\k$ at places where $\psi=0$ and $\bar\k = \k'$ where $\psi=1$. The modified bending energy of a flat membrane~\cite{Shlomovitz2008},
\bea
H = \int dS\, \hf \left[ \k(1-\psi) + \k' \psi \right] \left( c-\psi \bar c\right)^2 . \nn
\eea
For a spherical membrane with intrinsic curvature $c_s$, replacing $c$ by $2c_m$ where $c_m$ is the local mean curvature, the expression modifies to
\begin{align}
& H =  \int dS\, \hf \left[ \k(1-\psi) + \k' \psi \right] \left( 2c_m - c_s -\psi \bar c\right)^2\nn \\
& \approx \hf \k \int dS\, (2c_m - c_s)^2 - \k \bar c \int ds\, (2c_m - c_s)\psi \nn\\
&+ \hf \k \bar c^2 \int ds\, \psi^2. \nn
\end{align}
In the last line we kept terms up to bilinear order in curvature and AP concentration. This essentially means retaining only the change in spontaneous curvature due to AP concentration in the energy cost, and neglecting the impact of change in bending rigidity.  
In the main text we use $\bar H = -\bar c$, and $\L=\k \bar c^2$. 

\section{Quadratic expansion of bending energy}
\label{appendixA}
 Let  {\bf n } denotes the normal to the surface defined by $\Phi \equiv r(\theta,\phi,t)-r_0 u(\theta,\phi,t)=r_0$. 
 The sum of principal curvatures is proportional to the three-dimensional divergence of the normal to the surface, 
 $2 c_m = \nabla_3 \cdot {\bf n}$, 
 where the expression of the unit normal is,
\bea
{\bf n} = \f{\nabla_3 \Phi}{|\nabla_3 \Phi|} 
=\frac{\hat r -r_0\frac{\partial u}{r\partial \theta}\hat \h-r_0\frac{\partial u}{r \sin(\theta)\partial \phi} \hat\phi }{\left[1+\left(r_0\frac{\partial u}{r\partial \theta}\right)^2+\left(r_0\frac{\partial u}{r \sin(\theta)\partial \phi}\right)^2\right]^{1/2}}  \nn
\label{unitnormal}
\eea 
where  $\hat r$, $\hat \h$, $\hat \phi$ are the unit vectors in the $r$, $\theta$, $\phi$ directions respectively.
The total bending energy is therefore expressed as~\cite{Helfrich1986},
\bea
F_b=\frac{1}{2}\kappa\int (\nabla_3 \cdot {\bf n})^2\, ds-\kappa\int c_s{\nabla_3 \cdot \bf { n}}\, ds+\frac{1}{2}\kappa c_s^2\int ds, \nn
\label{totalbendingenergy2}
\eea
where,
\bea
\nabla_3 \cdot {\bf n}=\frac{2}{r_0}\left(1-u-\frac{1}{2}\Delta_2u+u^2+u\Delta_2u\right), \nn
\label{divn}
\eea
\bea
ds= \f{r^2}{n_r} d\Omega = r_0^2\left(1+2u+u^2+\frac{1}{2}(\nabla_2u)^2\right)d\Omega, 
\label{dS}
\eea
with $n_r$ the radial component of the normal ${\bf n}$. 
In the above relations $\nabla_2$ denotes the two-dimensional gradient on the surface of a sphere and $\Delta_2$ denotes the corresponding Laplace-Beltrami operator, given by 
\bea
\nabla_2 &=& \left(\frac{\partial}{\partial \theta},\frac{1}{\sin \theta }\frac{\partial}{\partial \phi}\right)\nn\\
\Delta_2 &=& \frac{1}{\sin \theta }\frac{\partial}{\partial \theta}\left(\sin \theta \frac{\partial}{\partial \theta}\right)+\frac{1}{\sin^2 \theta}\frac{\partial^2}{\partial \phi^2}. \nn
\label{nabla2}
\eea
Using the above equations we can express the total bending energy in terms of the function $u(\Omega,t)$ where $\Omega$ denotes a solid angle. Retaining terms up to second order in $u$
\bea
F_b &=& 2\kappa\int d\Omega\left[u^2+u\Delta_2u+\frac{1}{4}\left(\Delta_2u\right)^2\right] \nn\\
&=&  2\kappa\int d\Omega\left[ u + \f{1}{2} \Delta_2 u \right]^2
\label{bendingenergy3}
\eea
with $d\Omega = \sin\h d\h d\phi$.

\section{Equilibrium fluctuations and the flat membrane limit}
\label{equi_mem}
At equilibrium using $\delta F/\delta \psi=0$, from Eq.(\ref{tot_en}), one obtains
\beq
\psi = \frac{\k \bar H }{ \L r_0}\left(u+\hf \D_2 u\right),
\label{equin}
\eeq 
denoting non-uniform distribution of $\psi$ coupled with membrane deformation. 
At equilibrium, mean deformation $\la u \ra=0$ and mean AP concentration $\la \psi \ra=0$. To explore fluctuations,
we expand the fields into spherical harmonics $Y_{lm}(\h, \phi)$ (with $l = 1,\dots,\infty$, dropping $l=0$ term due to the volume-constraint as shown in the main text, and $m=-l,\dots,l$) such that 
$ u = \sum_{l,m} u_{lm}(t)  Y_{lm}(\h, \phi) $, $ \psi =  \sum_{l,m} \psi_{lm}(t) Y_{lm}(\h, \phi) $,  and use the eigenvalue equation $\D_2 Y_{lm} = -l(l+1) Y_{lm}$. Using Eq.(\ref{equin}) in the expression of free energy Eq.(\ref{tot_en}), and $\int d\Omega\, (\nabla_2 u)^2 = - \int  d\Omega\, u \D_2 u$, one can write
\begin{align}
&F =  \sum_{lm} \le[ \le(2\k -\f{\k^2 \bar H^2}{2\L}\ri) \le( \f{{\cal L}}{2} -1\ri)^2 + \s r_0^2 \le( \f{{\cal L}}{2} +1\ri) \right. \nn\\
&\left. + \f{K_t r_0^4}{2} \ri] |u_{lm}|^2  
\equiv \sum_{lm} \hf K_l |u_{lm}|^2, 
\label{eq_modes}
\end{align}
where ${\cal L}= l(l+1)$.
Eq.(\ref{eq_modes}) determines the equilibrium fluctuations $\la |u_{lm}|^2  \ra = \kb T/K_l$ through the equipartition theorem.  
The expression $\la |u_{lm}|^2  \ra = \kb T/K_l$ 
denotes the equilibrium deformation of the membrane due to the membrane-bound activator proteins. 
The corresponding non-uniform protein distribution is described by 
$\la | \psi_{lm} |^2 \ra = \f{\k^2 \bar H^2}{\L^2 r_0^2 } \left(\f{l(l+1)}{2} -1 \right)^2 \la | u_{lm}|^2 \ra $,
which is obtained by using Eq.(\ref{equin}). 
The associated pattern formation in the presence of a finite correlation length for density fluctuations of APs has captured supramolecular organization on the membrane of spherical bacteria {\em Staphylococcus aureus}~\cite{Garcia-Lara2015, Agudo-Canalejo2017}.    

 In the limit of large $l$ one obtains the flat membrane limit by replacing $l$ with $q r_0$, where $q$ denotes the wave number corresponding to a mode~\cite{Loubet2012}.
The above expression simplifies by using  $[l(l+1)/2 \,\pm 1] \approx l^2/2$ to give,
\beq
\la |u_q|^2 \ra r_0^4 = \f{\kb T}{\k_{\rm eff} q^4 + \s q^2}
\eeq
with a reduction of effective bending rigidity $\k_{\rm eff} = \k - \k^2 \bar H^2/4\L$, in agreement with flat membrane result in Ref.~\citenum{Ramaswamy2000}.

\section{Actin-myosin pull}
\label{appendixC}
The Greens function is given by $G_\a(\Omega, \Omega') = \sum_{lm} Y^\ast_{lm}(\Omega') Y_{lm}(\Omega) \exp[-\a l(l+1)]$. Here we find the result of the integral
$J = \int d\Omega' G_\a(\Omega, \Omega') f_p \varphi(\Omega')$. We use the expansion of $\varphi(\Omega') = \sum_{l' m'} \varphi_{l' m'}(t) Y_{l' m'} (\Omega')$. Using the 
orthonormality condition $\int Y^\ast_{lm}(\Omega')  Y_{l' m'} (\Omega') d\Omega' = \d_{l l'} \d_{m m'}$, we obtain 
\beq
\int d\Omega' G_\a(\Omega, \Omega') f_p \varphi(\Omega')= \sum_{l m } \varphi_{l m}(t) e^{-\a\, l(l+1)} Y_{lm} (\Omega)
\eeq
\section{Linearized dynamics}
\label{app:dyn}
Expanding the relative deformations $u$, AP concentration $\psi$, and myosin concentration  $\varphi$ in spherical harmonics with amplitudes $u_{lm} (t)$, $\psi_{lm} (t)$ and $\varphi_{lm} (t)$, respectively, the coupled linear differential equations describing the dynamics can be expressed as
\begin{eqnarray}
\frac{d u_{lm}}{d t} &=& -\G\left[ 4\kappa \le( \f{{\cal L}}{2} -1\ri)^2 + 2\s r_0^2 \le( \f{{\cal L}}{2} -1\ri) + K_t r_0^4 \right]u_{lm}  \label{udotlm}\nn\\ 
  &+& \le[ f_r - \G \k \bar H r_0 \le( \f{{\cal L}}{2} -1\ri) \ri] \psi_{lm} - f_p e^{-\a\,{\cal L}} \varphi_{lm},\label{udotlm}\\
\frac{d \psi_{lm}}{d t} &=&-D\,{\cal L}\, \psi_{lm} - \f{\mu \k \bar H r_0}{2} {\cal L}^2\left[1-\frac{2}{{\cal L}}\right]u_{lm},  \label{psidotlm}\\
\frac{d \varphi_{lm}}{d t} &=& -k_0 \varphi_{lm}+k_{\rm on} \psi_{lm}. \label{phidotlm}
\end{eqnarray}
where we used ${\cal L} := l(l+1)$, and the fact that the Laplace-Beltrami operator $\D_2$ is diagonal in the basis of spherical harmonics $\D_2 Y_{lm}(\h,\phi) = -l(l+1) Y_{lm}(\h,\phi)$.
In deriving the above relations, we used 
$\int d\Omega\, u^2 = \sum_{lm} |u_{lm}|^2$, 
$\int d\Omega\, \D_2 u = 0$, 
$\int d\Omega\, (\D_2 u)^2 = \sum_{lm} {\cal L}^2\,|u_{lm}|^2$,
$\int d\Omega\, u \D_2 u = -\sum_{lm} {\cal L} |u_{lm}|^2$.
The last term on the right hand side of Eq.(\ref{udotlm}) involving $\varphi_{lm}$, appears after performing the integration  
$\int d\Omega' G_\a(\Omega,\Omega')f_p \varphi(\Omega')$. Plugging in the expression of the 
 Greens function $G_\a(\Omega, \Omega') = \sum_{lm} Y^\ast_{lm}(\Omega') Y_{lm}(\Omega) \exp[-\a {\cal L}]$, along 
with the expansion $\varphi(\Omega',t) = \sum_{l' m'} \varphi_{l' m'}(t) Y_{l' m'} (\Omega')$ in the integral, and using the 
orthonormality condition $\int Y^\ast_{lm}(\Omega')  Y_{l' m'} (\Omega') d\Omega' = \d_{l l'} \d_{m m'}$, one obtains 
\beq
\int d\Omega' G_\a(\Omega, \Omega') f_p \varphi(\Omega')= \sum_{l m } f_p e^{-\a\, {\cal L}} \varphi_{l m}(t) Y_{lm} (\Omega) .
\eeq


\section{Stability matrix}
\label{app_m}
The elements of ${\bf S}^l$ are given by
$s^l_{11} = -4\tilde{\kappa}\Big[\frac{{\cal L}}{2}-1\Big]^2 -2\tilde{\sigma}\Big[\frac{{\cal L}}{2}+1\Big]- \tilde{K_t}$, 
$s^l_{12} = \tilde{f_r} - \tilde{\kappa}\tilde{H}\Big[\frac{{\cal L}}{2}-1\Big]$,
$s^l_{13} = -\tilde{f_p}\exp[-\alpha {\cal L}]$,
$s^l_{21} = -\frac{\tilde{\mu}\tilde{H}\tilde{\kappa}}{2}{\cal L}^2\Big[1-\frac{2}{{\cal L}}\Big]$,
$s^l_{22} = -{\cal L},~ s^l_{23} = 0,~ s^l_{31} = 0,~s^l_{32} = \tilde{k}_{on},~ s^l_{33} = -\tilde{k}_0$, where ${\cal L} := l(l+1)$. 
The linear stability of the system is determined by the nature of the eigenvalues of ${\bf S}^l$. The eigenvalue equation $\lambda^3+a_l\lambda^2+b_l\lambda+c_l  = 0$. The mathematical expressions of the coefficients are
\begin{eqnarray}
a_l &=& 4\tilde{\kappa}\left[\frac{{\cal L}}{2}-1\right]^2 +2\tilde{\sigma}\left[\frac{{\cal L}}{2}+1\right] +\tilde{K_t} +{\cal L}+\tilde{k}_0\nonumber\\ 
b_l&=&\left({\cal L}+\tilde{k}_0\right)\left[4\tilde{\kappa}\left(\frac{{\cal L}}{2}-1\right)^2 +2\tilde{\sigma}\Big[\frac{{\cal L}}{2}+1\Big] +\tilde{K_t} \right]\nonumber\\
&+&{\cal L}\tilde{k}_0+\frac{\tilde{\mu}\tilde{H}\tilde{\kappa}}{2}{\cal L}^2\left(1-\frac{2}{{\cal L}}\right)\left[\tilde{f_r} - \tilde{\kappa}\tilde{H}\left(\frac{{\cal L}}{2}-1\right)\right]\nonumber\\
c_l &=&{\cal L}\tilde{k}_0\left[4\tilde{\kappa}\left(\frac{{\cal L}}{2}-1\right)^2 +2\tilde{\sigma}\left(\frac{{\cal L}}{2}+1\right) + \tilde{K_t}\right]\nonumber\\
&-&\frac{\tilde{\mu}\tilde{H}\tilde{\kappa}}{2}\left[\left\{{\cal L}-1\right\}^2-1\right] \times \nn\\
&& \left[\tilde{f_p}\tilde{k}_{on}e^{-\alpha {\cal L}}-\tilde{k}_0\left\{\tilde{f_r}-\tilde{\kappa}\tilde{H}\left(\frac{{\cal L}}{2}-1\right)\right\}\right]. \nn\\
\label{eq:coeff_abc}
\end{eqnarray}

\section{Linear stability analysis involving a cubic polynomial}
\label{ap:linstab}
Here we present the details of calculating the generic phase diagram in Fig.~\ref{fig:phase_scheme}. 
We rewrite Eq.(\ref{eq:char}) as 
\begin{eqnarray}
p(\lambda) &=& \lambda^3+a_l\lambda^2+b_l\lambda\label{eq:cu1}\\
p(\lambda) &=& - c_l  \label{eq:cu2}
\end{eqnarray}
Intersections between Eq.(\ref{eq:cu1}) and Eq.(\ref{eq:cu2}) give real roots of Eq.(\ref{eq:char}). Eq.(\ref{eq:cu1}) is a cubic polynomial passing through the origin in $\lambda$-$p(\lambda)$ plane. As has been shown in Eq.(\ref{eq:coeff_abc}) $a_l > 0$. The cubic polynomial can have one minimum at $\l_m = -\frac{1}{3}a_l + \frac{1}{3}\sqrt{a_l^2-3b_l}$ and  one maximum at $\l_M=-\frac{1}{3}a_l - \frac{1}{3}\sqrt{a_l^2-3b_l}$ with $\l_M < \l_m$. Eq.(\ref{eq:cu2}) denotes a straight line parallel to $\lambda$ axis.

Eq.(\ref{eq:coeff_abc}) shows that although $a_l > 0$, $b_l$ and $c_l$ can change sign. We can construct a generalized stability diagram in $c_l$-$b_l$ plane as their signs hold the key  to the properties of Eq.(\ref{eq:cu1}) and Eq.(\ref{eq:cu2}) leading to various possible combinations of roots of Eq.(\ref{eq:char}), determining the nature of stability. 
The roots $\lambda_{1,2,3}$ come in the following combinations: ($i$)~$\lambda_{1,2,3}\textless 0$ give rise to a linearly stable ($s$) phase where the perturbation exponentially decays with time. The two cases ($ii$)~$\lambda_{1}\textgreater 0$ and $\lambda_{2,3}\textless 0$ and ($iii$)~$\lambda_{1,2}\textgreater 0$ and $\lambda_{3}\textless 0$ correspond to linearly unstable ($u$) phase as small perturbations grow exponentially with time. The case ($iv$)~$\lambda_{1}\textless 0$ and $\lambda_{2,3} = -\alpha\pm i\beta$, with real $\alpha$, $\beta$ denotes a (un-)stable spiral $ss$ ($us$) phase if $\a > 0$ ($\a<0$) as the amplitude of oscillation decays (grows) with time. 
Let us define $G(\lambda) := (\lambda-\lambda_1)(\lambda-\lambda_2)(\lambda-\lambda_3)$. The factor theorem gives $G(\l) = p(\l) + c_l$ leading to $c_l = -\l_1 \l_2 \l_3$, $b_l=(\l_1 \l_2 + \l_2 \l_3 + \l_3 \l_1)$, and $a_l = -( \l_1 + \l_2 + \l_3)$.

{\bf The boundary between linearly stable ($s$) and unstable ($u$) phase:} A transition from $s$- to $u$-phase takes place when among the three negative real roots one changes sign and becomes positive. The condition for the $s$-phase is satisfied when both $c_l,~b_l\textgreater~0$.
The instability appears ($u$-phase) if any one of the roots turns positive requiring $c_l~\textless 0$.
Thus the phase boundary between the $s$- to $u$-phase is given by,
\begin{equation}
c_l = 0. 
\label{eq:s_u_pb}
\end{equation}

The function $c_l$ depends on $l$ and Eq.(\ref{eq:s_u_pb}) denotes an $s$- to $u$-transition for a specific $l$- mode. The general condition for the onset of instability can be identified by setting the minimum of $c_l$ to zero. Minimizing $c_l$, using  the discretized condition $[ c_{l+1} - c_l ]_{l_{\rm min}}= 0$ 
numerically, we obtain an estimate of $l_{\rm min}$, to identify the $l$-independent phase boundary by setting $c_{l_{\rm min}} = 0$.

{\bf The boundary between linearly stable ($s$) and stable spiral ($ss$) phases:} A transition from $s$- to $ss$-phase happens when the line in Eq.(\ref{eq:cu2}) touches the minimum of the polynomial Eq.(\ref{eq:cu1}) in the negative side of the $\lambda$ axis in $\lambda$-$p(\lambda)$ plane making the complex conjugate roots to be real and degenerate at $\lambda_{m}$. As we compare the coefficients of the function $G(\lambda) := (\lambda-\lambda_1)(\lambda-\lambda_m)^2$ with $a_l$, $b_l$, and $c_l$ we end up with equations,
\begin{eqnarray}
a_l &=& -( \lambda_1+2\lambda_m ), \nonumber\\
b_l &=& \lambda_m(2\lambda_1+\lambda_m), \nonumber\\
c_l &=& - \lambda_1\lambda_m^2, \nonumber
\end{eqnarray}
where $\l_1$ is always negative. 
In the $s$-phase all eigenvalues are negative. Thus all coefficients $a_l,\, b_l,\, c_l$ are positive. Solving the first two equations for $\l_1$ and using it in the last equation, we obtain the phase  boundary between $s$- and $ss$-phase as,
\begin{equation}
c_l = \left(\frac{a_l}{3}+\frac{2}{3}\sqrt{a_l^2-3b_l}\right)\left(-\frac{a_l}{3}+\frac{1}{3}\sqrt{a_l^2-3b_l}\right)^2
\label{eq:s_ss_pb}
\end{equation}
with $b_l \geq 0$. 

{\bf The boundary between stable spiral ($ss$) and unstable spiral ($us$) phase:} The unstable spiral~($us$) phase is characterized by $\lambda_1\textless 0$ and $\lambda_{2,3}=g \pm i \, h$.
The $ss$- to $us$-transition takes place when $g$ changes sign. Consequently, the phase boundary is at $g=0$ leading to $\lambda_{2,3}=\pm i\, h$. Comparing the coefficients, $a_l$, $b_l$ and $c_l$ with $G(\lambda) := (\lambda-\lambda_1)(\lambda^2+h^2)$ we get conditions that read as,
\begin{eqnarray}
 a_l = - \lambda_1, ~
b_l = h^2, ~
c_l = - \l_1 h^2.  \nn
\end{eqnarray} 
These conditions lead us to the equation of the phase boundary,
\begin{equation}
c_l - a_lb_l = 0.
\label{eq:ss_us_pb}
\end{equation}
As $a_l > 0$, the above condition requires both $b_l > 0$ and $c_l > 0$, therefore the $ss$-$us$ phase boundary lies in the first quadrant of Fig.\ref{fig:phase_scheme}.

The shape fluctuation is determined by the superposition of all unstable $l$ modes. The behavior is controlled by $f_l:= c_l-a_lb_l$. 
In the presence of a small non-zero real part $g$ of $\l_{2,3}$, we can write $f_l \approx g\,(\l_1^2+a_l^2) + 2 g\, h^2$, i.e., $f_l \sim g$.
For $l$-modes giving $f_l > 0$ corresponding to $g>0$ the system shows $us$-phase. On the other hand, $f_l<0$~($g<0$)  corresponds to $ss$-phase. The phase boundary, as shown above, is at $f_l=0$. For a system in which $f_l$ varies non-monotonically with $l$ displaying a maximum, a band of $l$-modes obeying $f_l \geq 0$ shows $us$- behavior. This happens for parameter values, e.g., in Fig.~\ref{fig:phase_dia_sig_pus2}. In such a case, one can {\em maximize}  $f_l$ using the discretized condition $[f_{l+1} - f_l]_{l_{\rm max}} \approx 0$ to find $l_{\rm max}$. The condition $f_{l_{\rm max}} = 0$ gives the phase  boundary between $ss$ and $us$ phase, above which all $l$-modes are stable. 

{\bf The boundary between unstable spiral ($us$) and unstable ($u$) phase:} The transition between $us$  and a linearly unstable $u$ phase requires $b_l\textless 0$, as this transition,  is possible only if the minimum of the cubic polynomial lies at a positive $\lambda$ value (note that for $b_l=0, \lambda_m=0$). The instability appears as the minimum of the polynomial in Eq.(\ref{eq:cu1}) crosses Eq.(\ref{eq:cu2}). At that point, the two degenerate roots $\lambda_{2,3}$ become real positive while $\lambda_1$ is still real negative. The positive roots make the perturbation grow in time exponentially and the system becomes linearly unstable.

The boundary between $us$ and $u$ phase in second quadrant of Fig.\ref{fig:phase_scheme} can be calculated as we did for $s$- to $ss$-transition in the first quadrant and end up with the same mathematical expression as in Eq.(\ref{eq:s_ss_pb}), but now with $b_l\textless 0$. Eq.(\ref{eq:ss_us_pb}) and Eq.(\ref{eq:s_ss_pb}) meet at origin and this is the only point, where a direct transition from $s$ phase in the first quadrant to $u$-phase in the second quadrant possible.


\section{Stability of the mode $l=1$} 
\label{appendix_l0_l1}
For $l=1$ mode one can calculate the coefficients of the eigenvalue equation and get 
\begin{eqnarray}
a_1 &=& 4\tilde{\sigma} + \tilde{K_t} +\tilde{k}_0\nonumber\\ 
b_1 &=& (2+\tilde{k}_0)[4\tilde{\sigma} + \tilde{K_t}]+2\tilde{k}_0\nonumber\\ 
c_1 &=& 4\tilde{k}_0\left[2\tilde{\sigma} +\f{\tilde{K_t}}{2} \right]. 
\label{eq:l_1_mode}
\end{eqnarray}
These coefficients are always positive, limiting all the eigenvalues to be either real negative, or one real negative, and a pair of complex conjugates eigenvalues with a negative real part. Consequently, $l=1$ mode is either linearly stable or would show a decaying oscillation in time and thus eliminating any possibility of instability or pattern formation. 

\section{Transition from $ss$ to $us$}
\label{app_fl}
\begin{figure}[!t]
 \begin{center}
 \includegraphics[width=8.6cm]{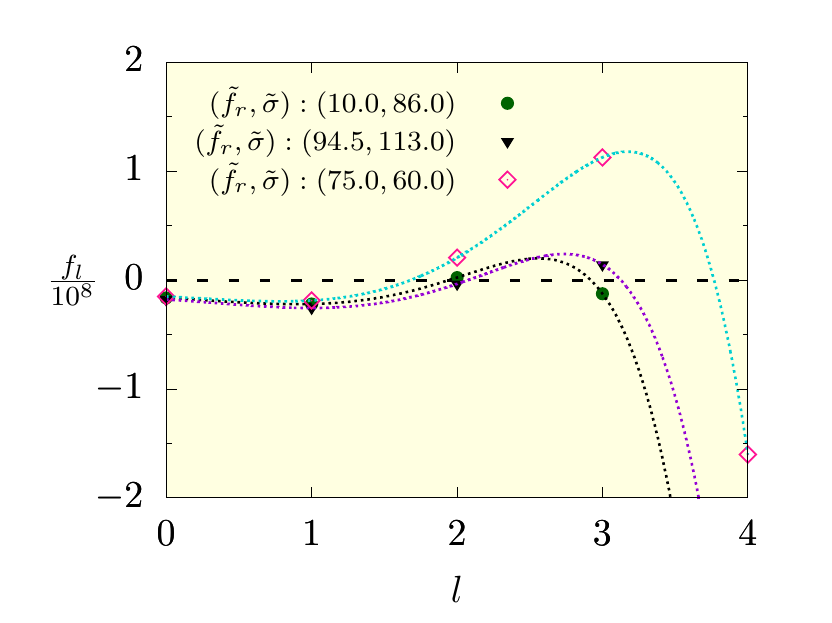}  
 \end{center}
 \caption{The function $f_l$ is plotted at three different parameter values shown in the figure. Other parameters are fixed at values $\tilde{H}=-1$, $\tilde{\mu} = 0.15$, $\tilde{K_t} = 10^3$, $\tilde f_p=50$, as in  Fig.~\ref{fig:phase_dia_sig_pus2}(${\bm a})$. Clearly $us$ phase characterized by $f_l>0$ appears for different $l$-values depending on the parameters $(\tilde f_r, \tilde \s)$.
}
 \label{fig:fl}
 \end{figure}

In Fig.~\ref{fig:fl} we show the variation of the function $f_l := c_l - a_l b_l$ as a function of $l$, at three sets of parameter values ($\tilde f_r, \tilde \s$), keeping all other parameter values same as in Fig.~\ref{fig:phase_dia_sig_pus2}(${\bm a})$. All of them show a non-monotonic variation. The range of $l$-values for which $f_l \geq 0$ shows $us$ behavior  depends on the dynamical parameters. Fig.~\ref{fig:fl} shows $us$ behavior for $l=2$ mode at $(\tilde f_r, \tilde \s)=(10,86)$, and for $l=3$ mode at $(\tilde f_r, \tilde \s)=(94.5,113)$. For the parameter combination $(\tilde f_r, \tilde \s)=(75,60)$, both the $l=2,\,3$ modes show $us$ as $f_l>0$ for both of them. 

\section{Running pulsation}
\label{app_shapes}
\begin{figure}[!t]
 \begin{center}
 \includegraphics[width=8.6cm]{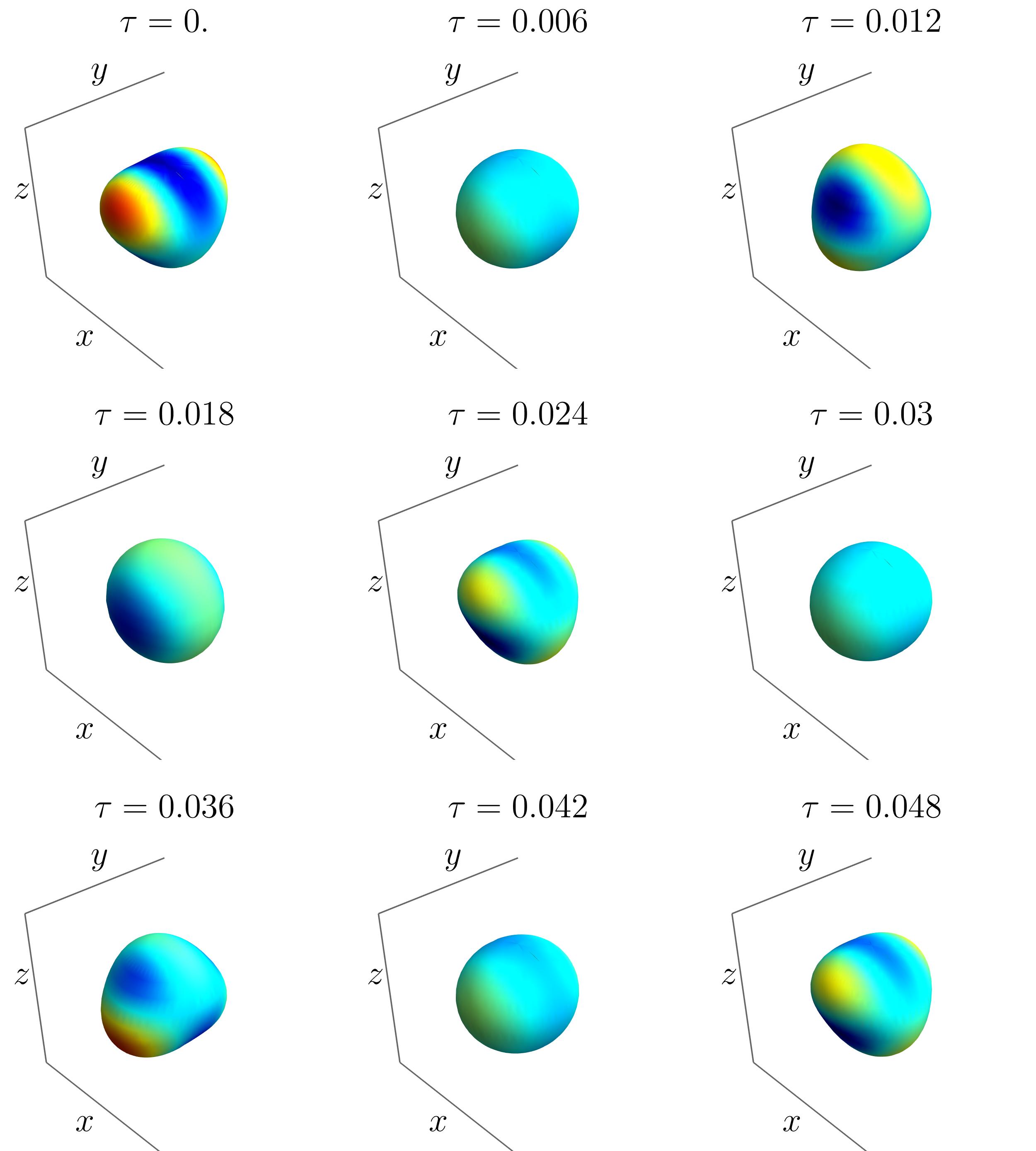}  
 \includegraphics[width=6cm]{bar_leg_dir_cur_l23.jpeg}
  \end{center}
 \caption{Plots showing the traveling wave generated on spherical membrane due to the superposition of $l=2$ and $3$ modes with nearly equal amplitude. 
 The color code on the deforming spherical shapes denotes  the local AP concentration $\psi$.
 The parameter values are $\tilde{H}=-1$, $\tilde{\mu} = 0.15$, $\tilde{K_t} = 10^3$, $\tilde f_p=50$,  $\tilde{\sigma} = 60$ and $\tilde{f_r} = 75$, as in Fig.~\ref{fig:kymo_comp_trav}.  The snapshots here correspond to the movie run\_puls.avi in the ESI. 
}
 \label{fig:curr_snap}
 \end{figure}

In Fig.\ref{fig:curr_snap} we show changes in the spherical shape as a pulsation runs between the two poles of the sphere. Associated with such deformations, as the movie run\_puls.avi in the ESI shows, the spherical membrane performs forward and backward somersaults with time.   

\section*{Author contributions}
DC designed the study with help from S Gutti. S Ghosh performed the linear stability analysis, numerical calculations, and data analysis under the supervision of DC. DC wrote the paper with assistance from S Ghosh. 

\section*{Conflicts of interest}
There are no conflicts to declare.

\section*{Acknowledgments}
We thank Gijsje Koenderink and Bidisha Sinha for valuable discussions and Thomas Litschel for bringing several relevant experiments to our notice. D.C. thanks SERB, India for financial support through grant number MTR/2019/000750, and International Centre for Theoretical Sciences for an associateship. 
SG acknowledges Debsuvra Ghosh for help with plots, and
thanks QuantiXLie Centre of Excellence, a project co-financed by the Croatian Government and European Union through the European Regional Development Fund - the Competitiveness and Cohesion Operational Programme (Grant No. KK.01.1.1.01.0004).

\bibliographystyle{prsty} 


\newpage
\onecolumngrid

\setcounter{figure}{0}

\section*{Electronic Supplementary Information (ESI) for \\
``Pattern formation, localized and running pulsation on active spherical membranes"}

\maketitle

 \section*{Additional Phase Diagrams and time evolutions}
 \label{appendix_phase_dia}
\begin{figure}[!h]
 \begin{center}
 \includegraphics[width=8cm]{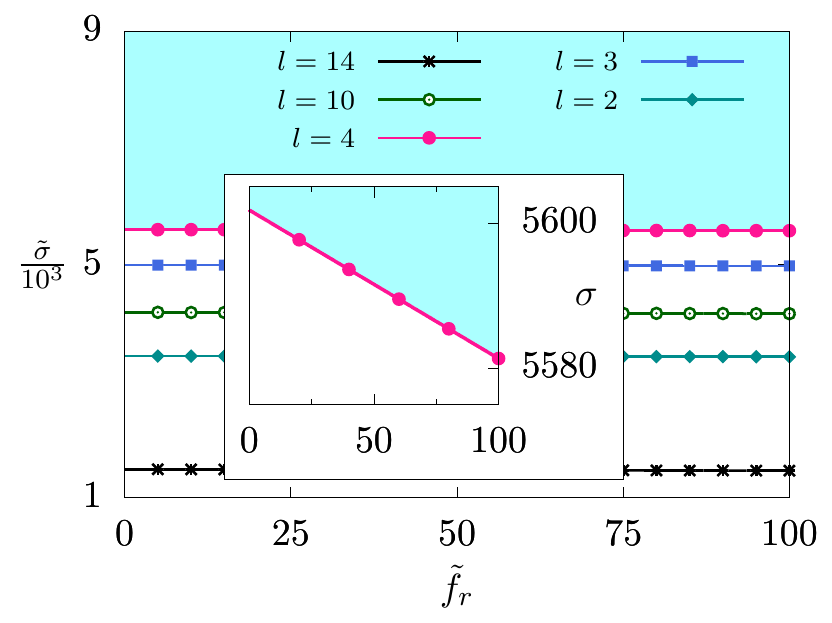}
 \end{center}
 \caption{Phase diagram for spherical membrane, using valley preferring APs with $\tilde{H}\textgreater~0$ in $\tilde{\sigma}-\tilde{f_p}$ plane showing dynamical transition form linearly stable ($s$) to unstable ($u$) phase  indicating pattern formation. As the actomyosin pull gets coupled with inward curvature producing APs, it brings about linear instability in the system through a positive feedback. Parameter values are, $\tilde{H} = 1$, and a fixed 
 $\tilde{f}_p = 10$, $\tilde{\mu} = 0.02$, $\tilde{K_t} = 2000$. All other parameters are fixed as in Table-1 of the main text. Each line corresponds to a $s$-$u$ boundary for a specific $l$-mode as indicated in the figure legend. In the region above these lines the system is stable corresponding to the particular $l$-mode. 
 The shaded region is linearly stable for all $l$ modes. }
 \label{fig:phase_dia_sig_pus1}
 \end{figure}

All the simulations are performed at a fixed bending modulus $\tilde \k = 25$, attachment rate $\tilde k_{on} = 3\times 10^4$, detachment rate $\tilde k_0 = 10$, and spread parameter of myosin pull $\a=0.001$. In Fig.\ref{fig:phase_dia_sig_pus1} in this ESI, we explore the impact of changing active parameter $\tilde{f}_r$ and the passive stabilizing factor of surface tension $\tilde \s$ keeping $\tilde{f}_p$ fixed.  Due to the positive $\tilde H=1$,  clustering of APs pull in the membrane locally. The active reaction $\tilde f_r$ due to F-actin polymerization, on the other hand, pushes the membrane outward. The surface tension $\tilde \s$ acts as a stabilizing factor.

\begin{figure}[!h]
 \begin{center}
 \includegraphics[width=12cm]{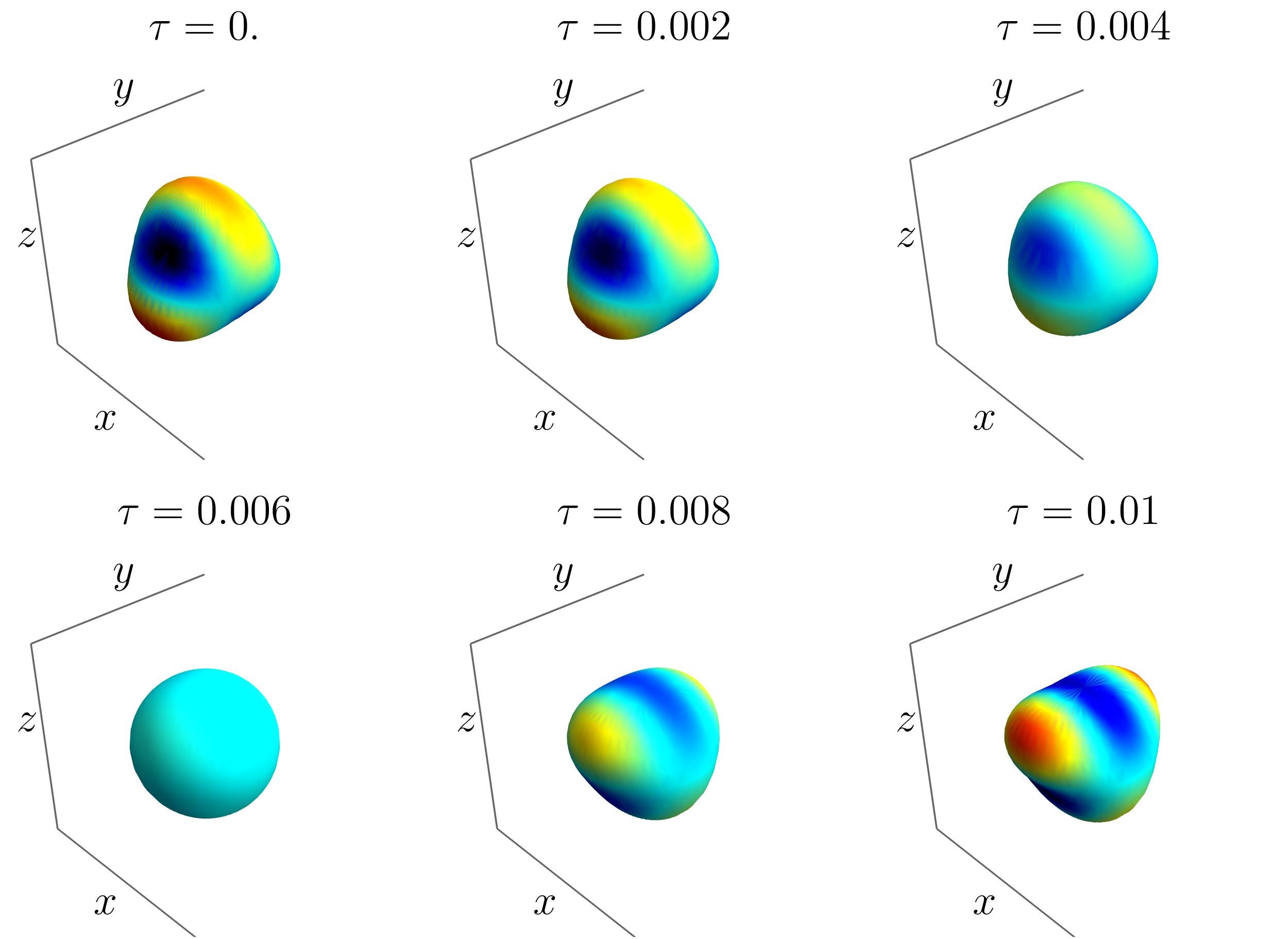}
 \includegraphics[width=6cm, angle=0]{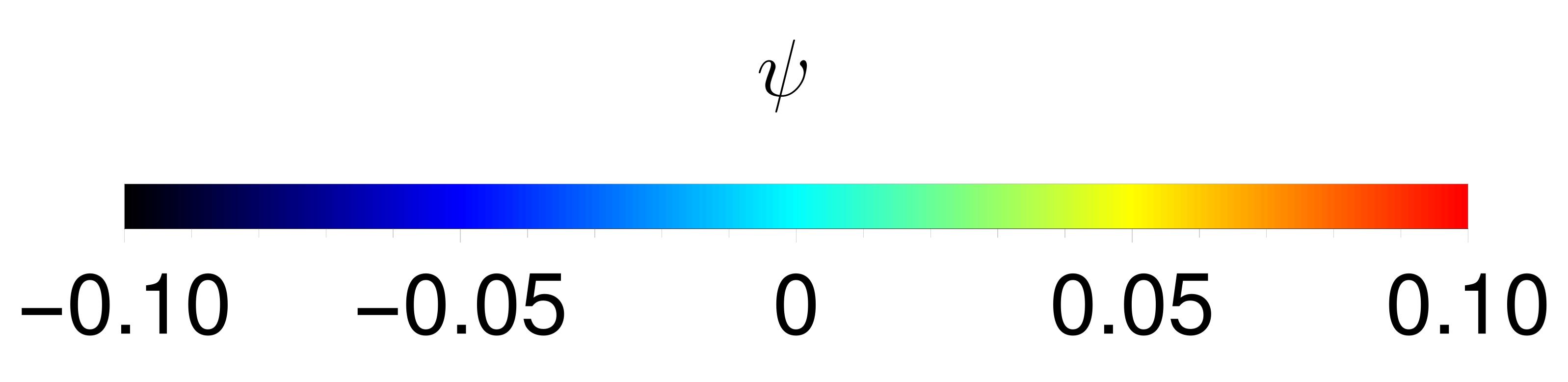} 
 \end{center}
 \caption{Plots showing localized pulsations corresponding to $l=3$ mode for a spherical membrane in which the  APs prefer hills.  The color code on the deforming spherical shapes denotes  the local AP concentration $\psi$. The parameter values used are 
 $\tilde H = -1$, $\tilde \mu=0.15$, $\tilde{K_t} = 1000$, $\tilde \s=113$, $\tilde f_r=94.5$, $\tilde{f_p} = 50$. }
 \label{fig:time_evo_os1}
 \end{figure}

\begin{figure}[!t]
 \begin{center}
 \includegraphics[width=12cm]{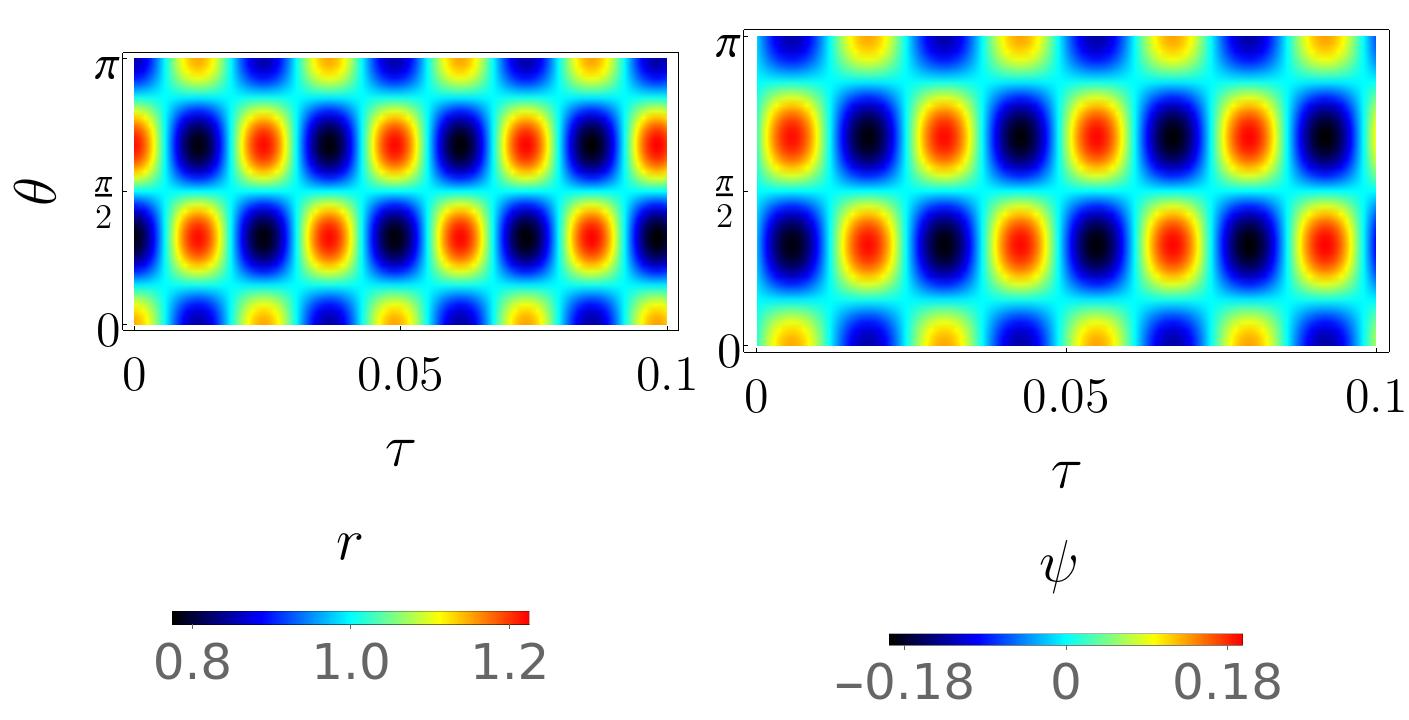}
 \end{center}
 \caption{Kymographs depicting the time evolution over azimuthal angle $\theta$ at $\phi = \frac{\pi}{2}$. The parameter values used are   $\tilde H = -1$, $\tilde \mu=0.15$, $\tilde{K_t} = 1000$, $\tilde \s=113$, $\tilde f_r=94.5$, $\tilde{f_p} = 50$. }
 \label{fig:kymo_comp}
 \end{figure}

Fig.~\ref{fig:time_evo_os1} of ESI shows localized pulsation corresponding to the excitation of $l=3$ mode at $\tilde f_r=94.5$, $\tilde \s=113$ corresponding to region ($iii$) of Fig.3(${\bm a}$) in the main text.
The corresponding kymograph in Fig.~\ref{fig:kymo_comp} of ESI shows the evolution of $u$ and $\psi$ along the polar angle $\h$ at a fixed $\phi=\pi/2$. 

\begin{figure}[!h]
 \begin{center}
 \includegraphics[width=12cm]{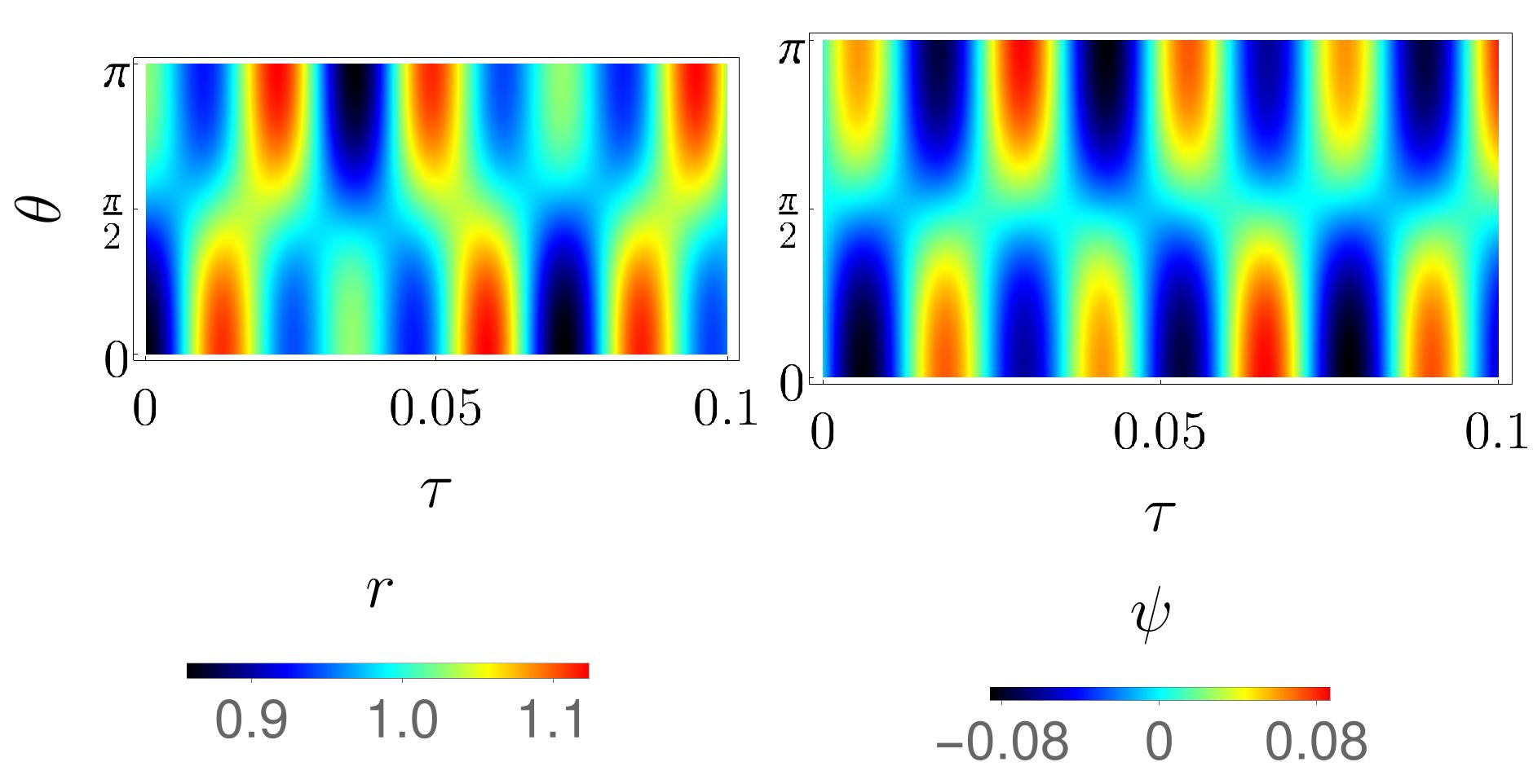}
 \end{center}
 \caption{Kymographs of running pulsation corresponds to Fig.~7 of main text. It shows evolutions of membrane deformation $r=r_0(1+u)$ and change in AP concentration $\psi$ over the polar angle $\theta$ at a fixed $\phi = \pi$. A running pulsation is generated with superposition of $l = 2,~3$ modes. The parameter values used here are $\tilde{H}=-1$, $\tilde{\mu} = 0.15$, $\tilde{K_t} = 10^3$, $\tilde f_p=50$,  $\tilde{\sigma} = 60$ and $\tilde{f_r} = 75$  corresponding to Fig.~7 and 11 of the main text. The connected nature of deformations is due to the traveling wave, and their slopes with respect to time indicate the velocities.}
 \label{fig:kymo_comp_trav_1}
 \end{figure}

In the main text, we have shown a kymograph of running pulsation  at $\phi=\pi/2$ in Fig.~7. In Fig.~\ref{fig:kymo_comp_trav_1} of ESI, we complement it with the kymograph at $\phi=\pi$ corresponding to the same set of parameter values, displaying the connected running pulsations. Along with Fig.~7 of  the main text, Fig.~\ref{fig:kymo_comp_trav_1} shows that the  local flux corresponding to the traveling wave depends on the location $(\h,\phi)$.

\begin{figure}[!h]
 \begin{center}
 \includegraphics[width=10cm]{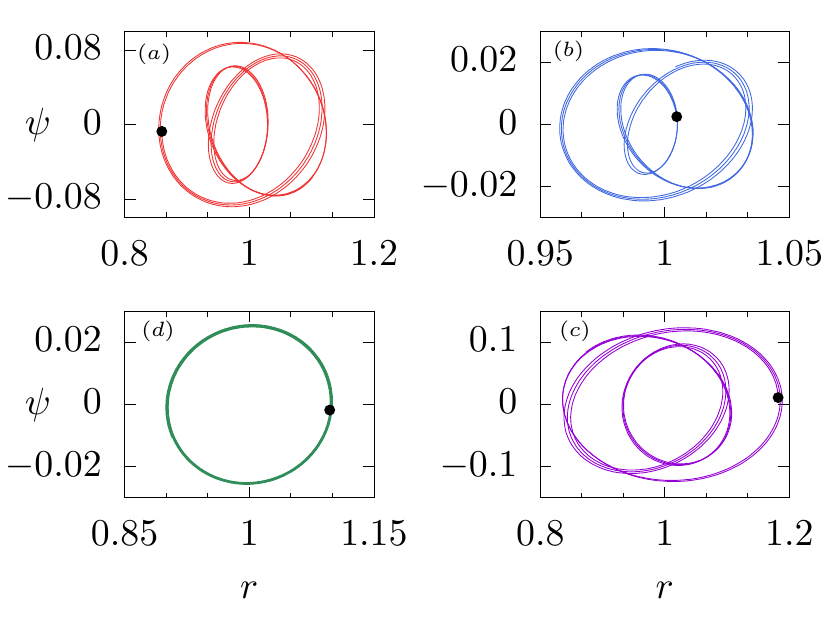}
 \end{center}
 \caption{Parametric plots of the field variables $r=r_0(1+u)$ and $\psi$ showing oscillations with multiple periodicity as the system displays traveling waves on the membrane surface. The plots display the behavior at azimuthal angles ($a$) $\theta = 0$, ($b$) $\theta = \frac{\pi}{6}$, ($c$) $\theta = \frac{\pi}{3}$ and ($d$) $\theta = \frac{\pi}{2}$. The parameter values used are  $\tilde{H}=-1$, $\tilde{\mu} = 0.15$, $\tilde{K_t} = 10^3$, $\tilde f_p=50$,  $\tilde{\sigma} = 60$ and $\tilde{f_r} = 75$, as in Fig.~7 of the main text. The filled black $\Circle$ in each plot indicates the initial state.}
 \label{fig:phase_plt_dir_cur}
 \end{figure}

Fig.~\ref{fig:phase_plt_dir_cur} of ESI shows parametric plots comprising of deformation field $u$ and local AP density field $\psi$ corresponding to running pulsation at a fixed set of $\theta$ and $\phi$ locations on top of the sphere. The parameter values are listed in figure caption, and correspond to Fig.7 in the main text. We note the existence of multiple periodicity at different locations, and the difference in amplitude of the oscillations in $r$ and $\psi$ at  various ($\theta$,$\phi$) points. 

\begin{figure}[!t]
\begin{center}
\includegraphics[width=10cm]{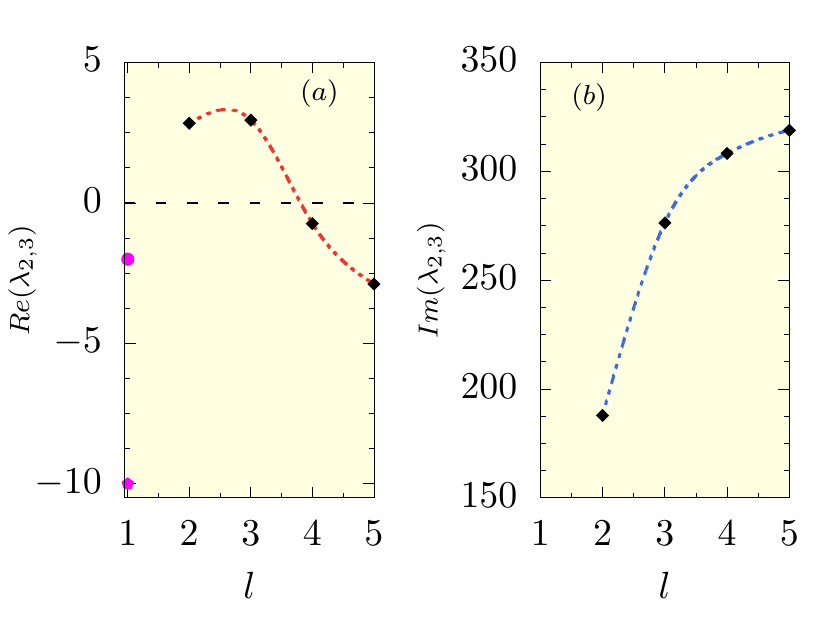}
\end{center}
\caption{The eigenvalues corresponding to running pulsation in Fig.~7 of main text. The parameter values used are $\tilde{H}=-1$, $\tilde{\mu} = 0.15$, $\tilde{K_t} = 10^3$, $\tilde f_p=50$,  $\tilde{\sigma} = 60$ and $\tilde{f_r} = 75$.  $\l_{2,3}$ are purely real and negative for $l=1$ and are denoted by the filled symbols $\pentagon$ ($\l_2$) and $\ocircle$ ($\l_3$). They have complex conjugate values with real part shown in ($a$) and the imaginary part in ($b$) using the filled $\Diamond$ symbol. $\l_1 < 0$ for all $l$.}
\label{fig:eigen_var_l_70_30}
\end{figure}

The $l$-mode dependence of real and imaginary parts of eigenvalues $\l_{2,3}$ distinguish  the nature of stable, linearly unstable, and unstable spiral phases.  This is shown in Fig.~\ref{fig:eigen_var_l_70_30} of ESI,  corresponding to the running pulsation at parameter values corresponding to Fig.s~7 and 11 of the main text.  

\begin{figure}[!h]
 \begin{center}
 \includegraphics[width=12cm]{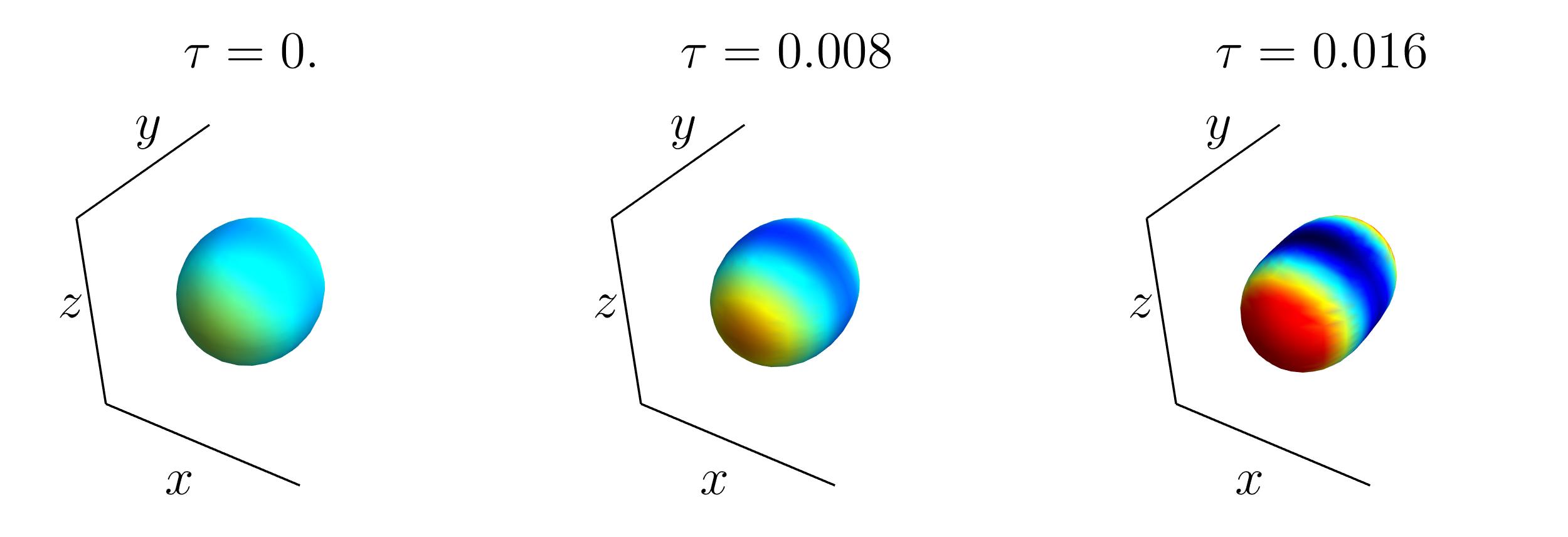}
  \includegraphics[width=6cm, angle=0]{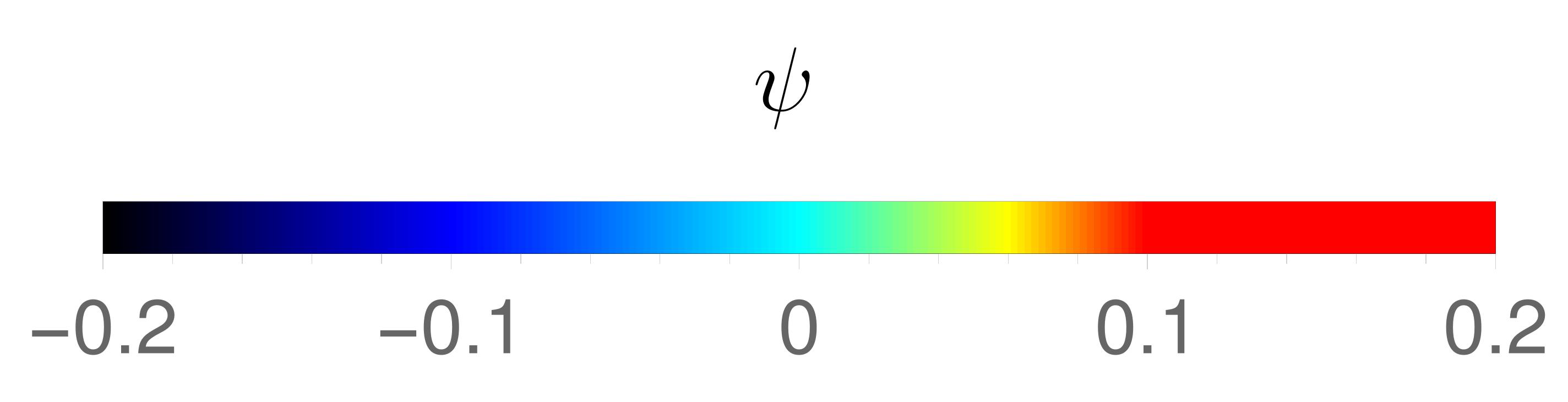}
   \end{center}
 \caption{Plots showing pattern formation on a spherical membrane due to linear instability in $l=2$ mode with $\tilde{H} = 1$, i.e., when the APs prefer local valleys. The color code on the deforming spherical shapes denotes  the local AP concentration $\psi$. The parameter values  used are $\tilde{\mu} = 0.02$, $\tilde{K_t} = 2000$, 
 $\tilde{f}_r = 10$, as in Fig.8 of main text, with $\tilde{\sigma} = 530$, $\tilde{f_p} = 5.0$.}
 \label{fig:time_evo_unstab_l2}
 \end{figure}

Finally in Fig.~\ref{fig:time_evo_unstab_l2} of ESI we depict conformational changes due to linear instability corresponding to the phase diagram Fig.8 of the main text, at $\tilde{\sigma} = 530$, $\tilde{f_p} = 5.0$, where $l=2$ mode is unstable. The corresponding evolution is qualitatively similar to the deformations of a cell at cytokinesis.

\section*{Choice of parameters}
\label{sec_para}
The cell can change its size and shape by regulating the osmotic pressure and effective surface tension. The bare surface tension $\s$ of such membranes can be $\sim 1\,$pN/$\mu m$~\cite{Betz2012}. This is tunable, e.g., the surface tension can be reduced incorporating more cholesterol in the membrane.

The actin polymerization rate at barbed end is $\sim 7.4 \mu{\rm M}^{-1}{\rm s}^{-1}$. With changing actin concentration from zero to 4\,$\mu$M it varies between $\approx 2 - 30$ subunit/s~\cite{Kuhn2005}. Considering subunit size $\sim 2.76$\,nm the actin growth rate gives a velocity $v_g=\,$5.52 to 82.8 nm/s.   
To directly translate it to protrusion of spherical membrane of radius $r_0 = 10\,\mu$m, the  relative growth rate $f_r=v_g/r_0$ is $\sim 5\times 10^{-4} -8\times 10^{-3}$\,s$^{-1}$. A myosin drive of F-actin with velocity $v_p \sim 1\,\mu$m/s~\cite{Kron1986}, leads to  $f_p=v_p/r_0\sim 0.1\,{\rm s}^{-1}$.

The viscosity of cytoplasmic extract is $\eta\sim 10\,$mPa-s~\cite{Valentine2005}. the viscous friction coefficient turns out to be $\g\approx 3 \pi \eta \xi \approx 2.8 \times 10^{-9}$\,N\,s/m$=2.8 \times 10^{-3}$\,pN-s/$\mu$m assuming the thickness of actin cortex $\xi\approx 30\,$nm~\cite{Clark2013}. Using the radius of spherical membrane $r_0=10\,\mu$m we get the membrane mobility coefficient 
$\G = 1/\g r_0^2 \approx 3.57 ~({\rm pN}\,\mu{\rm m\,s})^{-1}$.

The two dimensional diffusivity of AP  is $\sim 1\,\mu {\rm m}^2$/s~\cite{Shlomovitz2007, Chen2009a}. 
Considering a vesicle of radius $r_0 \approx 10\,\mu$m this gives the angular diffusivity  $D = 1\mu {\rm m}^2 {\rm s}^{-1}$/$r_0^2 = 10^{-2}$\,s$^{-1}$. 
Table-\ref{table_2} gives the list of parameter values used in the numerical calculations in this paper.

\begin{table*}
\small
  \begin{tabular*}{\textwidth}{@{\extracolsep{\fill}}lllll}
\hline
Parameters & Definition & Values & Scaled parameters & Scaled values\\
\hline \\
$D$ [s$^{-1}$] & angular diffusivity & $10^{-2}$ & unit & \\
$r_0 \,[\mu$m]& radius & 10 & unit & \\
$\G\,[({\rm pN}\,\mu{\rm m\,s})^{-1}$] & membrane mobility & 1.0& unit &\\
$\bar H\, [\mu {\rm m}^{-1}]$  & AP induced curvature &   $\pm 0.1 $~\cite{Shlomovitz2007} & $\tilde{H}=\bar{H}r_0$ & $\pm 1$ \\ 
$\mu  [({\rm pN}\,\mu{\rm m\,s})^{-1}$] & AP mobility & $0.5,\, 0.07$  &$\tilde{\mu} = \frac{\mu}{\Gamma}$ & $0.15$, $0.02$\\
$\s$ [pN$/\mu$m] & bare surface tension & 0 -- 2~\cite{Betz2012}& $\tilde{\sigma} = \frac{\sigma r_0^2\Gamma}{D}$ & 0 -- $2\times 10^4$\\
$\kappa$ [ $\kb T$] &  bending modulus & 60~\cite{Agudo-Canalejo2017} & $\tilde{\kappa} =\frac{\kappa\Gamma}{D}$& 25\\ 
$K_t$ [pN/$\mu$m$^3$] &  tether  & 0.001, 0.002~\cite{Alert2015} & $\tilde{K_t} =\frac{K_t r_0^4\Gamma}{D}$& 1000, 2000\\ 
$f_r\,[s^{-1}]$  & actin polymerization & 0 -- 1~\cite{Kuhn2005} & $\tilde{f_r} = \frac{f_r}{D}$ & 0 -- 100\\   
$f_p \,[s^{-1}]$ & myosin contraction &   0 -- 1 (${\cal O} [f_r]$ ) &$\tilde{f_p} = \frac{f_p}{D}$ & 0 -- 100 \\ 
$k_{\rm on}$\,[$s^{-1}$] & attachment rate   & $300$~\cite{Shlomovitz2007}& $\tilde{k}_{on} = \frac{k_{on}}{D}$ & $3\times 10^4$ \\
$k_0$\,[$s^{-1}$]  & bare detachment rate  & $0.1$~\cite{Shlomovitz2007} & $\tilde{k}_0 = \frac{k_0}{D}$ & $10$ \\
\hline
\end{tabular*}
\caption{The table lists all the parameters and their typical values used in the numerical calculations. $r_0$, $D$, and $\G$ set the units of length, time and force in the calculations. 
}
\label{table_2}
\end{table*}

\section*{Description of videos}
The three video in the ESI depict the time evolution in the three distinct dynamical regimes of the spherical membrane predicted by our model. The consecutive frames in localized pulsation (local\_pulse.avi), and the running pulsation (run\_puls.avi) are separated by dimensionless  time gaps $\D\tau = 10^{-4}$ and $\D\t = 2 \times 10^{-4}$, respectively. In the main text, snapshots from these movies are depicted in  Fig.s~5,\,6 and Fig.s~7,\,11 respectively.  The figure captions mention the values of the parameters used.

On the other hand, the consecutive frames in the movie of pattern formation due to linear instability (pattern.avi) is separated by $\D\tau = 5 \times 10^{-3}$. The snapshots from this movie and corresponding parameter values are shown in Fig.9 of the main text.

\bibliographystyle{prsty}

\end{document}